\documentclass[fleqn,usenatbib]{mnras}

\usepackage{newtxtext,newtxmath}

\usepackage[T1]{fontenc}

\DeclareRobustCommand{\VAN}[3]{#2}
\let\VANthebibliography\thebibliography
\def\thebibliography{\DeclareRobustCommand{\VAN}[3]{##3}\VANthebibliography}

\usepackage{graphicx}	
\usepackage{amsmath}	
\usepackage{amsfonts}
\usepackage{amscd}
\usepackage{natbib}
\usepackage[flushleft]{threeparttable}
\usepackage{orcidlink}
\usepackage{booktabs}
\usepackage{multirow}

\newcommand{\kms}{\,\mathrm{km} \, \mathrm{s}^{-1}}
\newcommand{\gpcm}{\, \mathrm{g} \, \mathrm{cm}^{-3}}

\newcommand{\Msolpc}{\, \mathrm{M}_{\sun} \, \mathrm{pc}^{-2}}

\newcommand{\ar}{a_{\mathrm{R}}}

\newcommand{\kappaP}{\kappa_{\mathrm{P}}}
\newcommand{\kappaPUV}{\kappa_{\mathrm{P},\mathrm{UV}}}
\newcommand{\kappaPIR}{\kappa_{\mathrm{P},\mathrm{IR}}}
\newcommand{\kappaRUV}{\kappa_{\mathrm{R},\mathrm{UV}}}
\newcommand{\kappaRIR}{\kappa_{\mathrm{R},\mathrm{IR}}}
\newcommand{\alphavir}{\alpha_{\mathrm{vir}}}
\newcommand{\cs}{c_{\mathrm{s}}}

\newcommand{\kappaSem}{\kappa_{\mathrm{Sem}}}
\newcommand{\Mcloud}{M_{\mathrm{cloud}}}
\newcommand{\Rcloud}{R_{\mathrm{cloud}}}
\newcommand{\Sigmacloud}{\Sigma_{\mathrm{cloud}}}
\newcommand{\Msun}{\mathrm{M}_{\sun}}

\newcommand{\Mout}{\dot{M}_{\mathrm{out}}}
\newcommand{\vout}{v_{\mathrm{out}}}

\newcommand{\ftrap}{f_{\mathrm{trap}}}

\def\HII{H~{\sc ii} }
\defcitealias{Skinner_2015}{SO15}
\defcitealias{Tsang_2018}{TM18}
\defcitealias{Crocker_2018b}{C18}
\defcitealias{Krumholz_Thompson_2012}{KT12}
\defcitealias{Menon_2022b}{Paper I}
\defcitealias{Levy_2021}{L21}

\title[Radiation Pressure-driven outflows]{Outflows Driven by Direct and Reprocessed Radiation Pressure in Massive Star Clusters}

\author[S.~H.~Menon et al.]{
Shyam H.~Menon$^{\orcidlink{0000-0001-5944-291X}}$,$^{1}$\thanks{E-mail: shyam.menon@anu.edu.au (SHM)}
Christoph Federrath$^{\orcidlink{0000-0002-0706-2306}}$,$^{1,2}$\thanks{E-mail: christoph.federrath@anu.edu.au (CF)}
Mark R.~Krumholz$^{\orcidlink{0000-0003-3893-854X}}$,$^{1,2}$
\\
$^{1}$Research School of Astronomy and Astrophysics, Australian National University, Canberra, ACT~2611, Australia\\
$^{2}$ARC Centre of Excellence for Astronomy in Three Dimensions (ASTRO3D), Canberra, ACT~2611, Australia
}

\date{Accepted XXX. Received YYY; in original form ZZZ}

\pubyear{2023}

\begin{document}
\label{firstpage}
\pagerange{\pageref{firstpage}--\pageref{lastpage}}
\maketitle

\begin{abstract}
We use three-dimensional radiation hydrodynamic (RHD) simulations to study the formation of massive star clusters under the combined effects of direct ultraviolet (UV) and dust-reprocessed infrared (IR) radiation pressure. We explore a broad range of mass surface density $\Sigma \sim 10^2$--$10^5 \, \Msolpc$, spanning values typical of weakly star-forming galaxies to extreme systems such as clouds forming super-star clusters, where radiation pressure is expected to be the dominant feedback mechanism. We find that star formation can only be regulated by radiation pressure for $\Sigma \lesssim 10^3 \, \Msolpc$, but that clouds with $\Sigma \lesssim 10^5 \, \Msolpc$ become super-Eddington once high star formation efficiencies ($\sim 80 \%$) are reached, and therefore launch the remaining gas in a steady outflow. These outflows achieve mass-weighted radial velocities of $\sim 15$ -- $30\,\kms$, which is $\sim 0.5$ -- $2.0$ times the cloud escape speed. This suggests that radiation pressure is a strong candidate to explain recently observed molecular outflows found in young super-star clusters in nearby starburst galaxies. We quantify the relative importance of UV and IR radiation pressure in different regimes, and deduce that both are equally important for $\Sigma \sim 10^3 \, \Msolpc$, whereas clouds with higher (lower) density are increasingly dominated by the IR (UV) component. Comparison with control runs without either the UV or IR bands suggests that the outflows are primarily driven by the impulse provided by the UV component, while IR radiation has the effect of rendering a larger fraction of gas super-Eddington, and thereby increasing the outflow mass flux by a factor of $\sim 2$.
\end{abstract}




\section{Introduction}
\label{sec:Introduction}

Radiation pressure on dust grains is a potentially important mechanism in regulating star formation and disrupting dusty gas in star clusters \citep{Krumholz_Matzner_2009,Fall_2010,Murray_2010_clusters,Raskutti_2016,Thompson_Krumholz_2016,Raskutti_2017}, maintaining the vertical stability of starbursts and AGN discs \citep{Scoville_2003,Thompson_2005,Andrews_2011,Krumholz_Thompson_2012}, and launching winds from galaxies, Active Galactic Nuclei (AGN), star clusters, young massive stars, and evolved asymptotic giant branch (AGB) stars  \citep{Murray_2011,Roth_2012,Krumholz_2013,Davis_2014,Thompson_2015,Rosen_2016,Wibking_2018,Zhang_2018, Costa_2018, Hofner_2018}. In the context of star/cluster formation, radiation pressure provides a crucial contribution in the expansion of feedback-driven \HII regions/bubbles \citep{Draine_2011,Kim_2016}, which limit the integrated star formation efficiency ($\epsilon_*$) of giant molecular clouds (GMCs) and lead to their inferred short lifetimes \citep{Chevance_2020,Chevance_2022, Chevance_2022b}

The mechanism of radiation pressure operates through the absorption of momentum in photons by dust grains, and coupling this momentum to the gas through collisions. Photons in two broad frequency bands are relevant here: the direct UV/optical photons from young stars, and the dust-reprocessed IR photons. The opacity of dust grains to the former is typically $\kappa_{\mathrm{UV}} \sim 100$--$1000 \, \mathrm{cm}^{2} \, \mathrm{g}^{-1}$, and thereby clouds with surface densities $\Sigma \gtrsim \kappa_{\mathrm{UV}}^{-1} \sim 10$--$100 \, \Msolpc$ are optically thick to these photons, and therefore susceptible to dispersal by direct radiation pressure. That being said, the thermal pressure of photoionised gas can be comparable to or larger than radiation pressure in some range of $\Sigma$; indeed, semi-analytic models \citep{Krumholz_Matzner_2009,Fall_2010, Murray_2010_clusters,Kim_2016,Rahner_2017}, numerical simulations \citep{Kim_2018}, and observations \citep{Lopez_2011,Lopez_2014,Barnes_2020,Olivier_2021} find that radiation pressure is the dominant feedback mechanism only for clouds whose escape velocities are $\gtrsim 10 \, \kms$. Such conditions are realised in GMCs that go on to form young massive star clusters \citep{Zwart_2010}. On the other hand, the IR opacities of dust are significantly lower \citep[$\kappa_{\mathrm{IR}} \lesssim 10 \, \mathrm{cm}^2 \, \mathrm{g}^{-1}$; ][]{Semenov_2003}, and therefore require much higher cloud surface densities ($\Sigma \gtrsim 10^3 \, \Msolpc$) to effectively absorb these photons. However, if this condition is satisfied, IR photons can undergo repeated cycles of absorption and emission, enhancing the imparted momentum over the stellar UV/optical photon momentum \citep{Thompson_2005,Murray_2010_clusters}. This is the so-called \textit{multiple-scattering} regime, to differentiate it from the single-scattering regime, where the dust is optically thin to IR photons. Environments in the multiple-scattering regime in the local universe are primarily found in extreme regions such as dwarf starbursts and ultra-luminous infrared galaxies (ULIRGs) like Arp~220, which are subject to high external pressures ($P/k_{\mathrm{B}} \gtrsim 10^8 \, \mathrm{K}\,\mathrm{cm}^{-3}$). These environments potentially host the formation sites of super-star clusters \citep[SSCs; e.g.,][]{Mcgrady_2005,Zwart_2010,Turner_2015,LindaSmith_2020}, and represent a dense mode of star formation that might have existed more commonly at high redshift. Observations suggest that these clusters form stars very efficiently and are mostly bound, with the role of stellar feedback on their formation and evolution largely uncertain \citep{Turner_2017,LindaSmith_2020, Emig_2020, Villas_2020,Costa_2021,Hao_2022}.

Recently, observations using the Atacama Large Millimeter/Submillimeter Array (ALMA) have managed to study the young, embedded phase of SSC formation at high resolution ($\sim 2 \,\mathrm{pc}$) in the nearby dwarf starburst NGC~253, shedding light on the properties of their natal GMCs \citep{Leroy_2018}, and the young stellar populations in them \citep{Mills_2021}. \citet{Levy_2021} conducted follow-up observations at even higher resolution ($\sim 0.5 \, \mathrm{pc}$) that have managed to probe the cluster-scale kinematics and feedback in these SSCs. Intriguingly, they find evidence of massive outflows from 3 of the 14 SSCs they characterise, with outflow velocities comparable to the SSC escape velocities, and outflowing masses that are a non-negligible fraction of the cloud/stellar mass. They consider multiple possible mechanisms that could drive these outflows, suggesting that either dust-reprocessed IR radiation pressure and/or stellar winds are the most likely; recent work by \citet{Lancaster_2021} casts doubts on the latter possibility. The former mechanism was raised in light of the semi-analytic model of \citet{Crocker_2018b}, which suggested that IR radiation pressure can drive outflows for $\Sigma \gtrsim 10^5 \, \Msolpc$ -- a condition satisfied by some of the detected SSCs with outflows. 

However, we recently presented 3D grey radiation hydrodynamic (RHD) simulations in \citet{Menon_2022b} (\citetalias{Menon_2022b} hereafter) -- which use temperature-dependent $\kappa_{\mathrm{IR}}$ and a state-of-the-art RHD scheme \citep{Menon_2022} -- showing that IR radiation pressure has minor dynamical impacts on clouds, irrespective of $\Sigma$. This is primarily because $\kappa_{\mathrm{IR}}$ is too low for radiation forces to compete with gravity (Eddington ratios, $f_{\mathrm{Edd}} \lesssim 0.4$) even at high $\Sigma$\footnote{\citet{Crocker_2018b} overestimate the impact of radiation forces as they use an analytical power-law approximation for $\kappa_{\mathrm{IR}}$, which significantly overestimates the opacities at higher temperatures.}, and because the efficiency of momentum transfer from radiation to gas is lowered by radiation-matter anticorrelation -- echoing the conclusions of earlier simulations that probed lower $\Sigma$ ranges \citep{Skinner_2015,Tsang_2018}. This raises concerns regarding the possibility that radiation pressure could drive the sort of winds seen in NGC~253. However, these simulations did not consider the contribution of radiation pressure in the UV band, and focused solely on the IR radiation pressure. The factor $\sim 100$ higher opacities in the UV could increase $f_{\mathrm{Edd}}$ significantly, especially in clouds in the lower ranges of the multiple-scattering limit\footnote{For very high $\Sigma$ clouds, the IR radiation force can be factors $\gtrsim 10$~times the corresponding UV one; in these cases, inclusion of the UV component is expected to have negligible effects; an expectation we confirm below.}. UV radiation pressure also has the attractive property that even if the cloud is globally sub-Eddington to this force, it can eject gas in sight lines that have lower $\Sigma$ set by turbulence \citep{Thompson_Krumholz_2016, Raskutti_2017}. Therefore, in this paper, we extend the simulations of \citetalias{Menon_2022b} to model the radiation field in both the UV and IR bands, thereby including the contribution of the direct and reprocessed radiation pressure, and test the dynamical impacts they have on SSC-forming clouds. We also quantify the relative impacts the two forces have at different cloud surface densities to constrain the dominant feedback mechanisms in different environments/conditions \citep[see, e.g., Fig. 12 in][]{Krumholz_2019}.

The paper is organised as follows: In Section~\ref{sec:Methods} we describe the equations solved in our simulations, the numerical prescriptions we use, and the initial conditions of our clouds. 
In Section~\ref{sec:Results} we present the results of our simulation suite, exploring the dynamical impacts of radiation pressure on our model clouds, properties of outflows driven, and the dominant forces (UV vs IR) driving these outflows. In Section~\ref{sec:summary} we provide a summary of our results, and discuss them in the context of the observed outflows in NGC~253.

\section{Methods}
\label{sec:Methods}
The simulation setup in this study largely follows that of \citetalias{Menon_2022b}; therefore, we summarize the salient features of our setup below and refer the reader to Section 2 of \citetalias{Menon_2022b} for further details.

\subsection{Equations solved}
\label{sec:Equations}
We solve the non-relativistic RHD equations in two grey bands that represent the stellar UV and dust-reprocessed IR bands respectively, self-consistently computing the reprocessing of the UV to the IR by dust. We use the mixed-frame formulation \citep{Mihalas_1982} in the RHD equations, retaining terms that are of leading order in all limiting regimes of RHD \citep[see, e.g.,][]{Krumholz_2007a}, given by\footnote{Note that we denote tensor contractions over a single index with dots (e.g., $\mathbfit{a} \cdot \mathbfit{b}$), tensor contractions over two indices by colons (e.g., \mathbfss{A}:\mathbfss{B}), and tensor products of vectors without an operator symbol (e.g., $\mathbfit{a}\mathbfit{b}$).}
\begin{gather}
	\frac{\partial \rho}{\partial t} + \nabla \cdot (\rho \textbf{v}) = 0 \label{eq:continuityeq} \\
	\frac{\partial (\rho\textbf{v})}{\partial t} + \nabla \cdot (\rho \textbf{v}\textbf{v}) = - \nabla P - \rho \nabla \Phi + \mathbfit{G}_{\mathrm{UV}} + \mathbfit{G}_{\mathrm{IR}} \label{eq:gasMom} \\
	\frac{\partial E_{r,\mathrm{UV}}}{\partial t} + \nabla \cdot \mathbfit{F}_{\mathrm{UV}} =  j_* - cG^0_{v,\mathrm{UV}} \label{eq:EradUV} \\
	\frac{\partial \mathbfit{F}_{\mathrm{UV}}}{\partial t} + \nabla \cdot (c^2\mathbfss{P}_{r,\mathrm{UV}}) =  -c^2\mathbfit{G}_{\mathrm{UV}} \label{eq:FradUV} \\
	\frac{\partial E_{r,\mathrm{IR}}}{\partial t} + \nabla \cdot \mathbfit{F}_{\mathrm{IR}} = \rho \kappaPUV c E_{r,\mathrm{UV}}   -cG^0_{v,\mathrm{IR}} \label{eq:EradIR} \\
	\frac{\partial \mathbfit{F}_{\mathrm{IR}}}{\partial t} + \nabla \cdot (c^2\mathbfss{P}_{r,\mathrm{IR}}) =  -c^2\mathbfit{G}_{\mathrm{IR}} \label{eq:FradIR} \\
	P = \cs^2 \rho \label{eq:Pgas},
\end{gather}
where,
\begin{equation}
    \label{eq:G0}
    \begin{aligned}
    G^0_{v,\lambda} =\;& \rho \kappaP(E_{r,\lambda} - j_{\lambda}) + \rho \left(\kappa_{\mathrm{R},\lambda} - 2\kappa_{\mathrm{P},\lambda} \right) \frac{\mathbfit{v} \cdot \mathbfit{F}_{\lambda}}{c^2} \\
    &+ \rho \left( \kappa_{\mathrm{P},\lambda} - \kappa_{\mathrm{R},\lambda} \right) \left[\frac{v^2}{c^2}E_{r,\lambda} + \frac{\mathbfit{v}\mathbfit{v}}{c^2} :\mathbfss{P}_{r,\lambda}\right],
\end{aligned}
\end{equation}
and
\begin{equation}
\label{eq:G}
\mathbfit{G}_{\lambda} = \rho \kappa_{\mathrm{R},\lambda} \frac{\mathbfit{F}_{\lambda}}{c} - \rho \kappa_{\mathrm{R},\lambda} E_{r,\lambda}\frac{\mathbfit{v}}{c}  - \kappa_{\mathrm{R},\lambda} \frac{\mathbfit{v}}{c} \cdot \mathbfss{P}_{\lambda},
\end{equation}
and $\lambda \equiv (\mathrm{UV}, \, \mathrm{IR})$ represent the band evolved in a corresponding equation. In the above equations $\rho$ is the mass density, $P$ the gas thermal pressure, $\mathbfit{v}$ the gas velocity, $\Phi$ the gravitational potential, $\mathbfss{I}$ the identity matrix, and $c$ the speed of light in vacuum. In the radiation moment equations (Equations~\ref{eq:EradUV} -- \ref{eq:FradIR}), $E_{r,\lambda}$ is the lab-frame radiation energy density, $\mathbfit{F}_{\lambda}$ the lab-frame radiation momentum density, $\mathbfss{P}_\lambda$ is the lab-frame radiation pressure tensor, $\kappa_{\rm P, \lambda}$ and $\kappa_{\rm R, \lambda}$ are the Planck and Rosseland mean opacities with avarages computed over the IR and UV bands\footnote{To be precise, by Planck and Rosseland means here we mean averages over the frequency band weighted by $B_\nu(T)$ and $\partial B_\nu/\partial T$, respectively, where $B_\nu(T)$ is the Planck function and $T$ is the radiation temperature.}. $j_*$ represents the direct UV radiation contribution rate from sink particles \citep{Federrath_2010_Sinks,Menon_2022b} and $j_{\lambda}$ represents the solid-angle integrated diffuse emission via dust grains in the given band. Equation~\ref{eq:Pgas} is the closure relation for the gas pressure, for which we assume an isothermal equation of state in our simulations, i.e., $P=\cs^2 \rho$, where $\cs$ is the thermal sound speed of the gas. The assumption of an isothermal equation of state does not considerably affect our results as the thermal pressure is subdominant over the radiation pressure in our simulations, and plays a minor role in the dynamics of our clouds. What would be affected by this assumption is the fragmentation on small scales. Heating by accretion feedback suppresses fragmentation \citep{OffnerEtal2009,Bate2009rad,KrumholzEtAl2016,FederrathEtAl2017,GuszejnovEtAl2018,MathewFederrath2020}, but our current simulations do not resolve these small-scale fragmentation processes anyway. What matters for the present simulations is the radiation output from massive stars and sub-clusters, which is modelled by sampling from a standard initial mass function on un-resolved scales, i.e., our sink particles represent small star clusters rather than individual stars \citepalias[for details, see][]{Menon_2022b}.

We pause to explain the radiation energy source terms on the right-hand side of Equations~\ref{eq:EradUV} and \ref{eq:EradIR}. In Equation~\ref{eq:EradUV}, the term $j_*$ represents the UV photons emitted by the sink particles. We set the diffuse emission term $j_{\mathrm{UV}} = 0$, as the dust does not re-emit in the UV. In Equation~\ref{eq:EradIR}, the first term represents the contribution to the dust-reprocessed IR radiation under the (very reasonable) assumption that all the energy the dust has absorbed from UV photons is instantly reprocessed into the IR\footnote{The timescale for this to occur is the thermal equilibration timescale of a dust grain that absorbs a photon ($t_{\mathrm{eq}}$). A rough estimate for $t_{\mathrm{eq}}$ for a grain of radius $a$ is the thermal energy of the grain $E_{\mathrm{th}} = 4/3 \pi a^3 \rho C_{\rho} T$ -- where $C_{\rho}$ is the specific heat of the grain, $\rho$ is its density, and $T$ is its temperature -- divided by the rate at which it radiates energy $L = 4 \pi a^2 \sigma_{\mathrm{SB}} T^4 Q$ where $Q$ is the quantum efficiency of the grain, which we can approximate in the small-grain limit as $Q \sim (hc/2 \pi a k_{\mathrm{B}}T)^{-2}$. Using these relations, and plugging in reasonable values of $\rho \sim 3 \, \gpcm$ and $C_{\rho} \sim 10^7 \, \mathrm{erg} \, \mathrm{g}^{-1} \, \mathrm{K}^{-1}$ gives $t_{\mathrm{eq}} \sim 30 \, \mathrm{yr} \, (T/10 \mathrm{K})^{-5} (a/1 \mu \mathrm{m})^{-1}$; this is orders of magnitude shorter than any relevant timescale for our problem. See \citet{Draine_2001} for a more accurate and detailed version of this calculation, which nonetheless yields a qualitatively identical conclusion.}. This treatment of the IR radiation field is more consistent than the approach in \citetalias{Menon_2022b} where IR photons are injected directly with a term analogous to $j_*$. For the diffuse emission, we set $j_{\mathrm{IR}} = \ar T^4$, where $\ar$ is the radiation constant, to represent the emission in the IR by dust grains. We also invoke the assumption of radiative equilibrium for the IR radiation -- i.e., the dust temperature is always equal to the radiation temperature $T_r = (E_{r,\mathrm{IR}}/\ar)^{1/4}$. This assumption is justified in Appendix~A of \citet{Krumholz_2013} considering the regime we are studying\footnote{We point out that although $T_r$ as defined here does not \textit{explicitly} include terms containing $E_{r,\mathrm{UV}}$, their contribution is ensured through its effect on $E_{r,\mathrm{IR}}$ as governed by Equation~\ref{eq:EradIR} In other words, the increase of dust/radiation temperature via the absorption of UV photons is captured within this definition.}. The combination of the aforementioned assumptions implies that the first term in the parentheses in Equation~\ref{eq:G0} for the IR band is zero, and therefore net heating or cooling from IR radiation arises purely due to mechanical contributions.

To close the equations above, we require a closure relation for the radiation pressure tensor. In both bands, we adopt the variable Eddington tensor (VET) closure
\begin{equation}
\label{eq:closure}
\mathbfss{P}_{r,\lambda} = \mathbfss{T}_{\lambda}E_{r,\lambda},
\end{equation}
where $\mathbfss{T}_{\lambda}$ is the Eddington Tensor for a given band. We use an Eddington tensor directly calculated from angular quadratures of the band specific intensity $I_{r,\lambda}(\hat{\mathbfit{n}}_k)$, using the relations
\begin{gather}
\label{eq:angularmoment0}
E_{r,\lambda}=\int d \Omega\, I_{r,\lambda}(\hat{\mathbfit{n}}_k)/c, \\
\mathbfss{P}_{r,\lambda}=\int d \Omega\, \hat{\mathbfit{n}}_k \hat{\mathbfit{n}}_k\,I_{r,\lambda}(\hat{\mathbfit{n}}_k)/c.
\label{eq:angularmoment2}
\end{gather}
$I_{r,\mathrm{UV}}$ and $I_{r,\mathrm{IR}}$ are calculated from formal solutions of the time-independent radiative transfer equations in the respective bands,
\begin{gather}
    \frac{\partial I_{r,\mathrm{UV}}}{\partial s} = \frac{j_*}{4\pi} - \rho \kappaRUV I_{r,\mathrm{UV}}  \label{eq:RTeqUV} \\ 
    \frac{\partial I_{r,\mathrm{IR}}}{\partial s} = \rho \kappaRIR \left[\frac{c j_{\mathrm{IR}}}{4 \pi} - I_{r,\mathrm{IR}}\right] \label{eq:RTeqIR}
\end{gather}
where the term $j_*/(4\pi)$ represents the photons from the isotropically emitting sink particle, and $j_{\mathrm{IR}}$ is the frequency-integrated reprocessed emission of the dust grain at the temperature $T_r$, which is also assumed to be directionally isotropic. We use the grey Rosseland-mean opacity, $\kappaRUV$ ($\kappaRIR$) in Equation~\ref{eq:RTeqUV} (~\ref{eq:RTeqIR}) to ensure consistency with the choice of flux-mean opacity we made in the radiation moment equations.

\subsection{Numerical methods}

\renewcommand{\arraystretch}{1.0}
\begin{table*}
\caption{Summary of our simulation suite and their initial condition parameters.}
\centering
\label{tab:Simulations}
\begin{threeparttable}
\begin{tabular}{l c c c c c c c c c}
\toprule
\multicolumn{1}{l}{Model}& \multicolumn{1}{c}{$M_{\mathrm{cloud}}$}& \multicolumn{1}{c}{$R_{\mathrm{cloud}}$}& \multicolumn{1}{c}{$\Sigma_{\mathrm{cloud}}$}& \multicolumn{1}{c}{$n_{\mathrm{cloud}}$}& \multicolumn{1}{c}{$\sigma_{v}$}& \multicolumn{1}{c}{$v_{\mathrm{esc}}$}& \multicolumn{1}{c}{$t_{\mathrm{ff}}$}& \multicolumn{1}{c}{$\mathrm{UV}$}& \multicolumn{1}{c}{$\mathrm{IR}$}\\
& [$10^6 \, \Msun$]& [pc]& [$\Msun \, \mathrm{pc}^{-2}$]& [$\mathrm{cm}^{-3}$]& [km/s]& [km/s]& [Myr]& & \\
\midrule
\texttt{S2UVIR} &$1.0$ &$31.5$ &$3.2$$ \times 10^{2}$ &$3.1$$ \times 10^{2}$ &$12$ &16 &3.0 &\checkmark &\checkmark\\
\texttt{S3UVIR} &$1.0$ &$10.0$ &$3.2$$ \times 10^{3}$ &$9.7$$ \times 10^{3}$ &$22$ &29 &0.5 &\checkmark &\checkmark\\
\texttt{S4UVIR} &$1.0$ &$3.2$ &$3.2$$ \times 10^{4}$ &$3.1$$ \times 10^{5}$ &$40$ &52 &0.09 &\checkmark &\checkmark\\
\texttt{S5UVIR} &$1.0$ &$1.0$ &$3.2$$ \times 10^{5}$ &$9.7$$ \times 10^{6}$ &$71$ &92 &0.02 &\checkmark &\checkmark\\
\texttt{S2UV} &$1.0$ &$31.5$ &$3.2$$ \times 10^{2}$ &$3.1$$ \times 10^{2}$ &$12$ &16 &3.0 &\checkmark &$\times$\\
\texttt{S3UV} &$1.0$ &$10.0$ &$3.2$$ \times 10^{3}$ &$9.7$$ \times 10^{3}$ &$22$ &29 &0.5 &\checkmark &$\times$\\
\texttt{S4UV} &$1.0$ &$3.2$ &$3.2$$ \times 10^{4}$ &$3.1$$ \times 10^{5}$ &$40$ &52 &0.09 &\checkmark &$\times$\\
\texttt{S3IR} &$1.0$ &$10.0$ &$3.2$$ \times 10^{3}$ &$9.7$$ \times 10^{3}$ &$22$ &29 &0.5 &$\times$ &\checkmark\\
\texttt{S4IR} &$1.0$ &$3.2$ &$3.2$$ \times 10^{4}$ &$3.1$$ \times 10^{5}$ &$40$ &52 &0.09 &$\times$ &\checkmark\\
\texttt{S5IR} &$1.0$ &$1.0$ &$3.2$$ \times 10^{5}$ &$9.7$$ \times 10^{6}$ &$71$ &92 &0.02 &$\times$ &\checkmark\\
\bottomrule
\end{tabular}
\begin{tablenotes}
\small
\item \textbf{Notes}: The row in bold denotes the fiducial simulation of our study. Columns in order indicate - Model: model name, $M_{\mathrm{cloud}}$: mass of cloud, $R_{\mathrm{cloud}}$: radius of cloud, $\Sigma_{\mathrm{cloud}}$: mass surface density of the cloud given by $\Sigma_{\mathrm{cloud}} = M_{\mathrm{cloud}}/(\pi R_{\mathrm{cloud}}^2)$, $n_{\mathrm{cloud}}$: number density of the cloud given by $n_{\mathrm{cloud}} = 3M_{\mathrm{cloud}}/(4 \pi R_{\mathrm{cloud}}^3m_{\mathrm{H}})$ where $m_{\mathrm{H}}$ is the mass of atomic hydrogen, $\sigma_{v}$: turbulent velocity dispersion of the cloud, $v_{\mathrm{esc}}$: escape velocity of the cloud, $t_{\mathrm{ff}}$: free-fall time of the cloud, $\mathrm{UV}$: UV band is on ($\checkmark$) or off ($\times$), $\mathrm{IR}$: IR band is on ($\checkmark$) or off ($\times$).
\end{tablenotes}
\end{threeparttable}
\end{table*}

\label{sec:NumMethod} 
The numerical methods used to solve the equations outlined in the previous section are identical to \citet{Menon_2022b}. We use the Variable Eddington Tensor-closed Transport on Adaptive Meshes (\texttt{VETTAM}; \citealt{Menon_2022}) method coupled to the \texttt{FLASH} magneto-hydrodynamics code \citep{Fryxell_2000,Dubey_2008} for our simulations. For the hydrodynamic updates, we use an explicit Godunov method in the split, five-wave HLL5R (approximate) Riemann solver \citep{Waagan_2011}. The Poisson equation for the self-gravity is solved using a multi-grid algorithm implemented in \texttt{FLASH} \citep{Ricker_2008}. Sink particles are used to follow the evolution of gas at unresolved scales, the formation of which is triggered when gas properties satisfy a series of conditions to test for collapse and star formation \citep{Federrath_2010_Sinks}. Gravitational interactions of sink particles with gas and other sinks are considered, and a second-order leapfrog integrator is used to advance the sink particles\citep{Federrath_2010_Sinks,FederrathBanerjeeSeifriedClarkKlessen2011}.

Sink particles in our simulations represent unresolved sub-clusters rather than individual stars. As in \citetalias{Menon_2022b}, we assume that these sub-clusters fully sample the initial mass function (IMF) of a young stellar population, and adopt an appropriate fixed light-to-mass ratio of $\langle L_*/M_*\rangle = 1.7 \times 10^3 \, \mathrm{erg} \, \mathrm{s}^{-1} \, \mathrm{g}^{-1}$, where $M_*$ is the mass of the radiating source. The UV radiation from sink particles is then included via the term $j_*$ in Equation~\ref{eq:EradUV}, given by 
\begin{equation}
    \label{eq:jstar}
    j_{*}(r)=\frac{L_{*}}{\left(2 \pi \sigma_{*}^{2}\right)^{3 / 2}} \exp \left(-\frac{r^{2}}{2 \sigma_{*}^{2}}\right), 
\end{equation}
where $L_* = M_*\langle L_*/M_*\rangle$, and $r$ is the radial distance of a grid cell from the sink particle. We adopt a value of $\sigma_* = 4 \Delta x_{\mathrm{min}}$, where $\Delta x_{\mathrm{min}}$ is the minimum cell size in the domain; we have shown in \citetalias{Menon_2022b} that the radiation forces are fairly insensitive to the choice of this parameter. 

The radiation moment equations in the UV (Equations~\ref{eq:EradUV} and~\ref{eq:FradUV}) and IR (Equations~\ref{eq:EradIR} and~\ref{eq:FradIR}) bands are operator-split from the hyrodynamic and gravity updates, and solved with an implicit Euler-backward temporal scheme \citep{Menon_2022}. We perform two radiation updates per hydrodynamic timestep: first for the UV band, then followed by the IR band, which uses the time-updated solution in the UV band as a source term (i.e., the first term on the RHS of Equation~\ref{eq:EradIR}) -- hence the scheme is fully implicit in the radiation quantities\footnote{An alternate approach to treat the coupled nature of the two bands is to solve Equations~(\ref{eq:EradUV})--(\ref{eq:FradIR}) together in one global, implicit update for both bands. However, we found that the resulting performance and accuracy with this approach was inferior to the one we adopt. This is likely due to the fact that in a global update, the coupling between UV and IR bands has to be treated \textit{internally} in the solution of the linear system, and thus the equality of the energy lost to the UV band and gained by the IR band is enforced only to the level imposed by the linear solver tolerance. By contrast, in our two-step process we can guarantee the equality of these quantities to machine precision. A subtle point worth noting here is that our adopted approach is possible only because the coupling between the bands is unidirectional in frequency space -- i.e., from UV to IR. For a system where this is not the case, a single, coupled update would be required.}. The time-independent radiative transfer equations (Eq.~\ref{eq:RTeqUV} and~\ref{eq:RTeqIR}) for obtaining the VET closure are obtained with a hybrid characteristics ray-tracing scheme \citep{Buntemeyer_2016}, and is computed prior to the radiation moment update for the respective band.

In \citetalias{Menon_2022b} we performed a series of tests with \texttt{VETTAM} to quantify the accuracy of our VET-based RHD scheme for IR radiation. Since here is the first time that \texttt{VETTAM} is utilised to model UV radiation pressure, we reproduce the results obtained with our scheme for the fiducial model of \citet{Raskutti_2016} in Appendix~\ref{sec:AppendixUVRP}. \citet{Kim_2017} simulated this model with their Adaptive Ray-Tracing (ART) method, based on the \texttt{HARM}$^2$ algorithm introduced by \citet{Rosen_2017}, to demonstrate that the $M_1$ method (used in \citealt{Raskutti_2016}) underestimates the (UV) radiation forces, and as a result, the net star formation efficiency ($\epsilon_*$) -- obtaining $\epsilon_* \sim 25 \%$ with the ART scheme as opposed to $\sim 42 \%$ in the \citet{Raskutti_2016} version. We find a value of $\epsilon_* \sim 28 \%$, which is closer to the ART result than the $M_1$, demonstrating that a moment method based on the VET closure can be of comparable accuracy to an ART scheme for modelling the dynamical effects of streaming radiation forces\footnote{It is important to point out however, that an ART scheme, while quite accurate for streaming radiation, would be unable to model reprocessed or diffuse radiation (i.e., the IR band).}.

\subsection{Initial conditions and parameters}

We initialise our simulations as a uniform spherical cloud with mass ($\Mcloud$) and radius $\Rcloud$, which together define a cloud mass density $\rho_{\mathrm{cloud}} = \Mcloud/[(4/3) \pi \Rcloud^3]$ and a mass surface density $\Sigmacloud = \Mcloud/(\pi \Rcloud^2)$. The clouds are placed in a lower-density ambient medium with $\rho = \rho_{\mathrm{cloud}}/100$ in pressure-equilibrium, achieved using a mass-scalar to represent cloud material (see Section~2.4 of \citetalias{Menon_2022b}). The domain size is fixed to $L= 4 \Rcloud$ to allow sufficient volume to track potentially expanding material due to feedback. Clouds are initialised with turbulent velocities that follow a power spectrum $E(k) \propto k^{-2}$ with a natural mixture of solenoidal and compressive modes \citep[appropriate for supersonic molecular-cloud turbulence; see e.g.,][]{HeyerBrunt2004,Federrath_2013} for $k/(2\pi/L) \in \left[2,64 \right]$, generated with the methods described in \citet{Federrath_2010}, and publicly available \citep{FederrathEtAl2022ascl}. The velocity dispersion $\sigma_v$ is set such that the virial parameter $\alphavir=2$ where $\alphavir$ is given by
\begin{equation}
    \alphavir = \frac{2E_{\mathrm{kin}}}{E_{\mathrm{grav}}} = \frac{5\Rcloud \sigma_v^2}{3G\Mcloud},
\end{equation}
where $E_{\mathrm{kin}} = (1/2) \Mcloud \sigma_v^2$ and $E_{\mathrm{grav}} = (3/5) G \Mcloud^2/\Rcloud$. The sound speed $\cs$ is set such that the sonic Mach number $\mathcal{M} = \sigma_v/\cs = 11.5$. Our choice of $\alphavir$ ensures the cloud is marginally bound in its initial state; we do not explore variations of $\alphavir$ here since we found relatively minor differences in the competition between radiation and gravity in \citetalias{Menon_2022b} (Section 3.2.3) with different $\alphavir$ . We also do not include magnetic fields in our simulations; we discuss in \citetalias{Menon_2022b} the caveats associated with this. The domain boundary conditions for the hydrodynamics are set to diode -- i.e., gas is allowed to flow out of the domain, but not allowed to enter it.

The opacity in the UV band is set to a constant value of $\kappaPUV = \kappaRUV = 1000 \, \mathrm{cm}^2 \, \mathrm{g}^{-1}$, consistent with typical estimates of the gray radiation pressure cross section per H atom to blackbody radiation peaking at UV wavelengths \citep[blackbody temperatures $\sim$ few $\times 10^4 \, \mathrm{K}$; ][]{Draine_2011}. The opacity in the IR band is kept identical to \citetalias{Menon_2022b}, i.e., a temperature- (and density-) dependent infrared opacity with $\kappaPIR = 0$ (due to radiative equilibrium) and $\kappaRIR = \kappaSem$, where $\kappaSem =  \kappaSem (\rho, T_{\mathrm{r}})$ is the \citet{Semenov_2003} opacity, calculated at the radiation temperature $T_{\mathrm{r}}$. The temperature dependence of the opacity in the IR is retained, which is crucial to accurately capture the dynamics of the clouds under reprocessed radiation pressure \citepalias{Menon_2022b}. The initial condition for the radiation is as follows: $E_{r,\mathrm{UV}} = \mathbfit{F}_{\mathrm{UV}} = 0$, and $E_{r,\mathrm{IR}} = \ar T_{\mathrm{r},0}^4$,  $\mathbfit{F}_{\mathrm{IR}} = 0$, where $T_{\mathrm{r},0} = 40 \, \mathrm{K}$ is the initial dust temperature in the cloud. We adopt Marshak boundary conditions for the radiation field \citep{Marshak_1958}, with boundary radiation temperatures of $T_{\mathrm{b},\mathrm{UV}} = 0$ and $T_{\mathrm{b},\mathrm{IR}} = T_{\mathrm{r},0}$ respectively. We also note that the boundary condition for the ray-tracer is kept consistent with these choices.

We note that we do not treat photoionization of gas by UV photons, and the corresponding thermal-pressure driven feedback on the clouds. However, in the regime we are exploring (high surface-density clouds with escape speeds $\gtrsim 10 \, \mathrm{km}/s$), radiation pressure forces have been shown to exceed ionized gas pressure, and dominate the dynamical evolution of clouds \citep{,Dale_2012,Kim_2016,Kim_2018}. 

\begin{figure*}
    \centering
    \includegraphics[width=0.9 \textwidth]{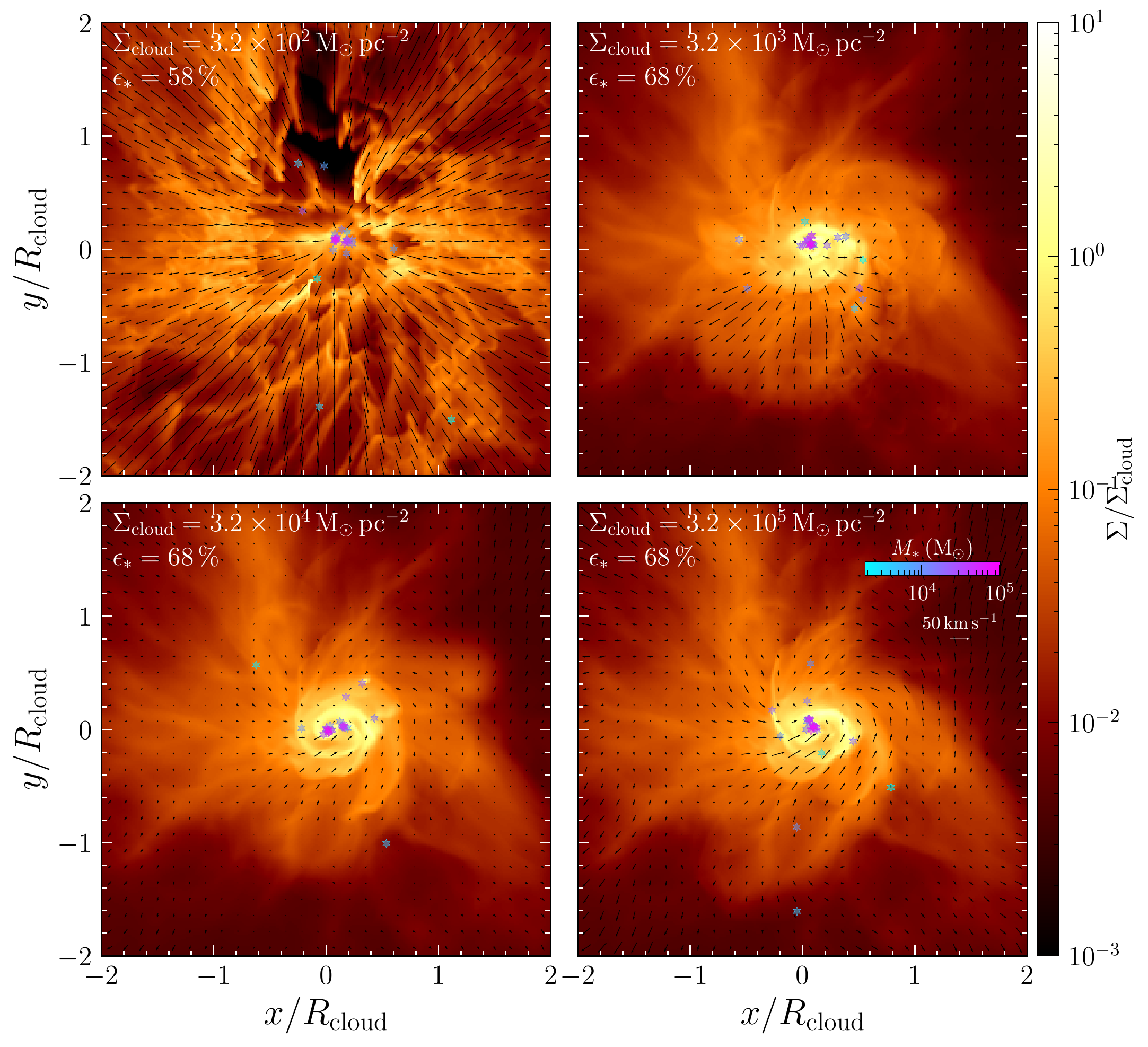}
    \caption{Surface density maps at $t = 3 t_{\mathrm{ff}}$ for the different values of $\Sigmacloud$ (panels)  with the corresponding star formation efficiency ($\epsilon_*$) annotated. Star symbols indicate sink particles, coloured by their mass (see inset colour bar in lower right panel). Vectors (black) indicate the mass-weighted projected velocity field, with arrow length indicating velocity magnitude. The scale for the velocity vectors is annotated in the lower right panel. The surface densities and positions are scaled to $\Sigmacloud$ and $\Rcloud$, respectively. Animations of the time evolution of these maps are available as supplementary online material.}
    \label{fig:CompareSigmasProj}
\end{figure*}

\begin{figure*}
    \centering
    \includegraphics[width=0.9 \textwidth]{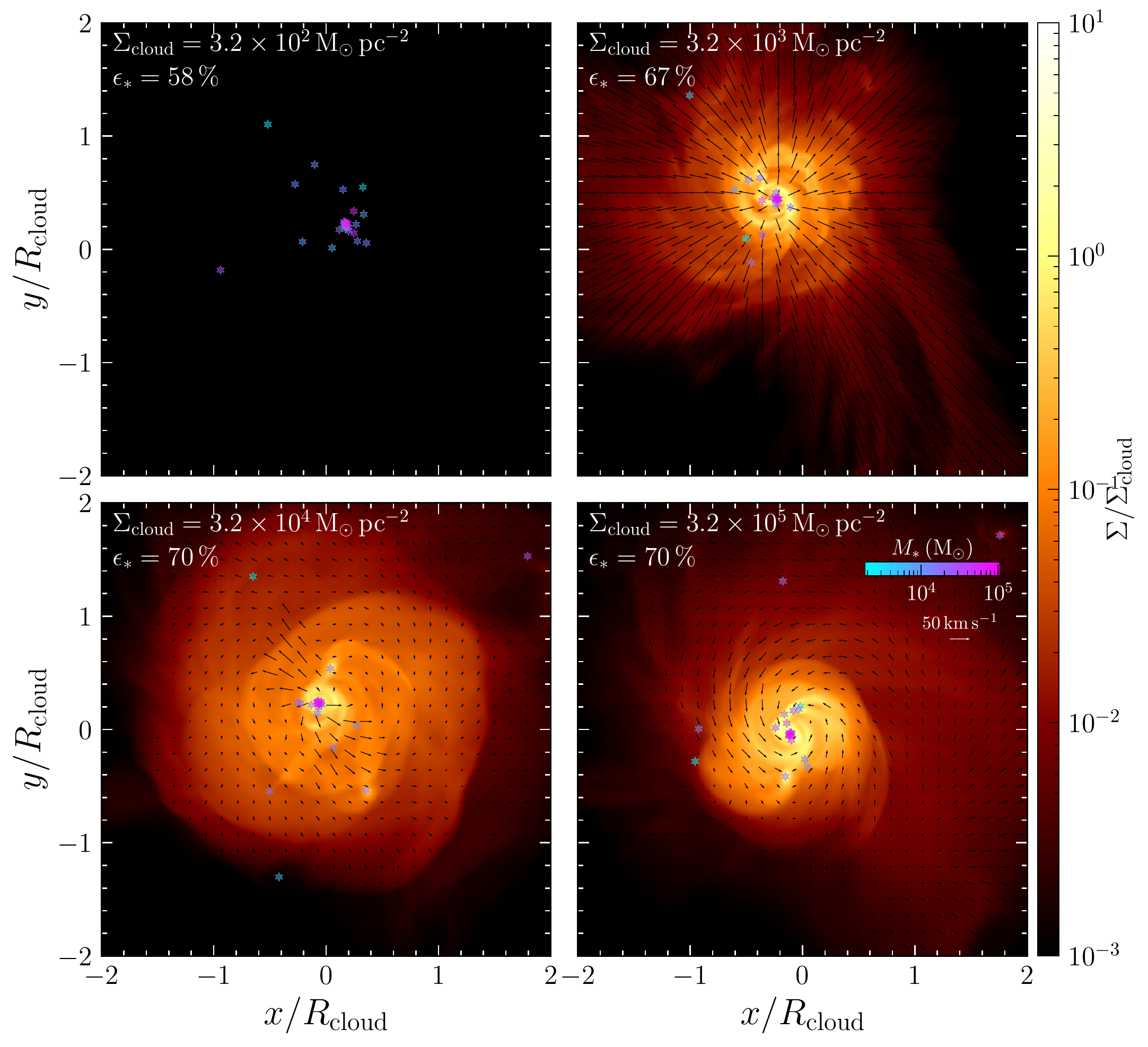}
    \caption{Same as Figure~\ref{fig:CompareSigmasProj}, but at time $t = 7 t_{\mathrm{ff}}$.}
    \label{fig:CompareSigmasProjLate}
\end{figure*}

\subsection{Simulations}
We run a range of simulations with different surface densities $\Sigmacloud$ -- along the lines of \citetalias{Menon_2022b} -- to test the impact of radiation pressure in different environments. We obtain our target values of $\Sigmacloud$ by keeping the mass of the clouds fixed to $\Mcloud = 10^6 \Msun$, and scaling $\Rcloud$ appropriately. We test values of $\Sigmacloud = 3.2 \times 10^2 \Msolpc$ up to $\Sigmacloud = 3.2 \times 10^5 \Msolpc$, varying by factors of 10 between consecutive runs with different $\Sigmacloud$; the resulting cloud parameters are tabulated in Table~\ref{tab:Simulations}. All of our clouds are optically thick to UV photons. We note that our parameters cover a range that is more massive and of higher surface density than typical star-forming clouds in local galaxies, a choice motivated by the expectation that radiation pressure is the dominant stellar feedback mechanism in this regime \citep{Krumholz_Matzner_2009,Fall_2010, Kim_2016}. The two lowest surface density points ($\Sigmacloud \sim 10^2$--$10^3 \, \Msolpc$) represent conditions appropriate for young massive clusters in regions like the Central Molecular Zone (CMZ), whereas the two higher values of $\Sigmacloud$ ($\Sigmacloud \sim 10^4$--$10^5 \, \Msolpc$) represent super-star clusters that are probably found only in more extreme environments such as starburst galaxies \citep[e.g.,][]{Leroy_2018}.

Our standard runs evolve radiation in both the UV and IR bands. To isolate the effects of the radiation pressure in either band, and to quantify their relative importance in the evolution of the clouds, we also run some control simulations where either the UV or IR band is not included. We list all the simulations explored in this study in Table~\ref{tab:Simulations}. The IR-only runs have already been presented in \citetalias{Menon_2022b}; the UV-only runs are new. We adopt as a convention that run names are of the form \texttt{SsUVIR}, \texttt{SsUV}, and \texttt{SsIR} respectively for UV+IR, UV-only, and IR-only runs, while \texttt{s} encodes the cloud surface density $\Sigmacloud = 3.2 \times 10^{\texttt{s}} \Msolpc$. We do not simulate a UV-only version for $\Sigmacloud = 3.2 \times 10^{5} \Msolpc$, as we expect UV to be unimportant compared to IR at these high surface densities; low-resolution tests confirm this is the case. Similarly, we do not run an IR-only version for our lowest surface density case ($\Sigmacloud = 3.2 \times 10^{3} \Msolpc$), as it is below the typical surface densities required to be optically thick to IR photons ($\Sigma \lesssim \kappa_{\mathrm{IR}}^{-1} \sim 10^3 \Msolpc$).

All our simulations use a uniform grid (UG) with $N^3 = 256^3$ grid cells; for our domain of size $L = 4 R_{\rm cloud}$, this corresponds to a resolution in terms of the number of grid cells per cloud radius of $R_{\rm cloud}/\Delta x = 64$. We show that our results are converged with numerical resolution in Appendix~\ref{sec:AppendixResolution}. We adopt a CFL number of 0.4, a relative tolerance of $10^{-8}$ for our implicit update of the radiation moment equations, and perform the solution to the time-independent transfer equation with 48~rays per cell using our ray-tracing scheme \citep[based on the Healpix algorithm;][]{BuntemeyerEtAl2016}. We run all simulations up to the point where all the mass has been accreted onto sink particles or expelled from the computational domain by radiation forces, or to a time $t=8\,t_{\mathrm{ff}}$, where $t_{\mathrm{ff}}$ is the free-fall time of the cloud -- whichever is earlier.

\section{Results}
\label{sec:Results}

Here we present the main results of our study, beginning with a broad overview of the qualitative outcomes in Section \ref{ssec:evolution}. We follow this up with a detailed examining of the radiatively-driven outflows we observe in Section \ref{sec:Outflows}, a comparison of the relative roles of the IR and UV radiation forces in Section \ref{sec:UVIRForces}, and a quantitative analysis of the (in)efficiency of radiation in regulating star formation in Section \ref {sec:radpres_efficiency}.

\subsection{Evolution of Clouds}
\label{ssec:evolution}

\begin{figure}
    \centering
    \includegraphics[width =  0.44 \textwidth]{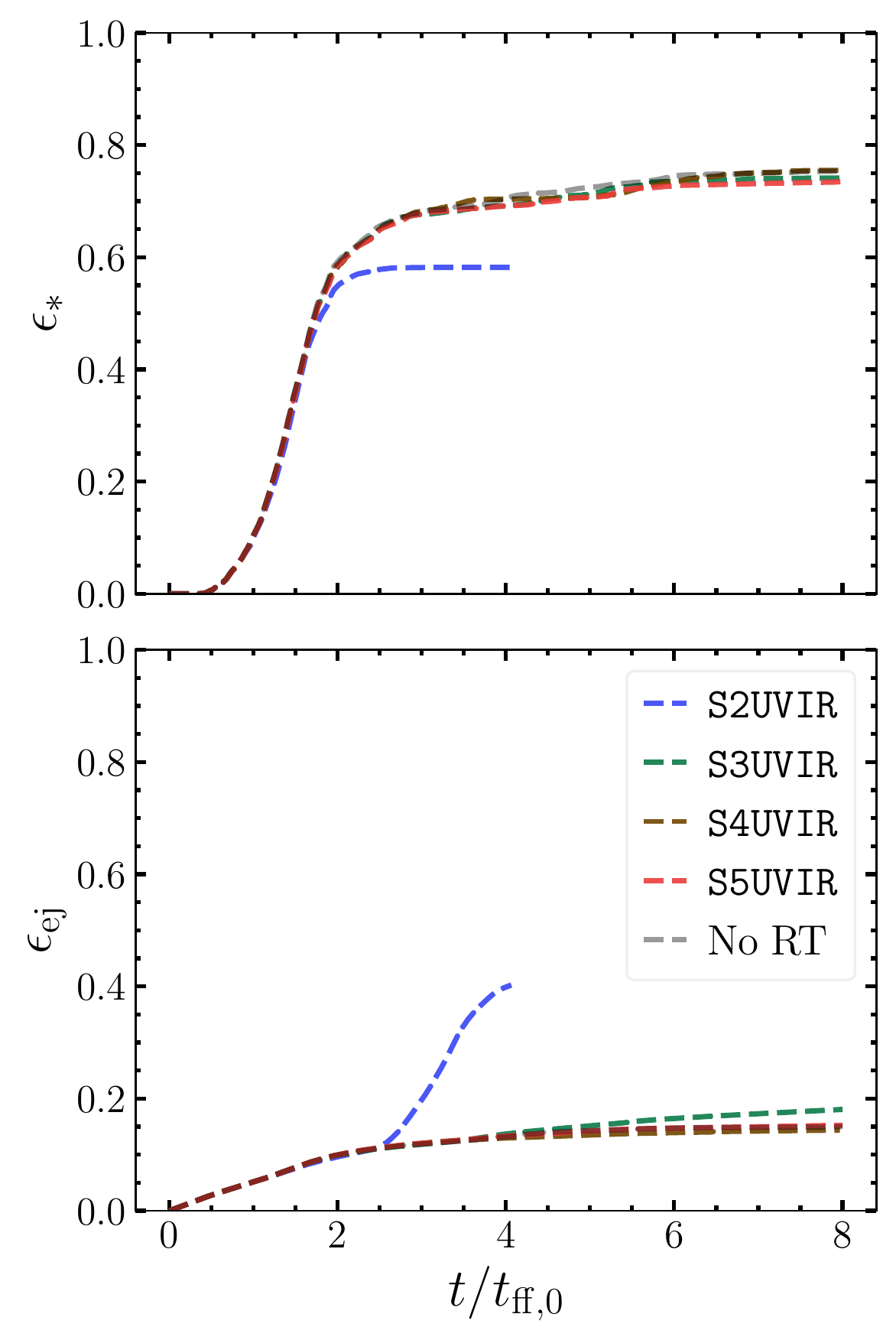}
    \caption{Time evolution of the integrated star formation efficiency ($\epsilon_* = M_*/\Mcloud$) (top panel) and the fraction of mass ejected from the computational volume ($\epsilon_{\mathrm{ej}} = M_{\mathrm{ej}}/\Mcloud$ where $M_{\mathrm{ej}}$ is the ejected mass) (bottom panel), for different values of $\Sigma_{\rm cloud}$ (colours). Dark grey dashed lines indicate a control run without radiation feedback (No RT).}
    \label{fig:CompareSFE}
\end{figure}

\begin{figure*}
    \centering
    \includegraphics[width =  0.94 \textwidth]{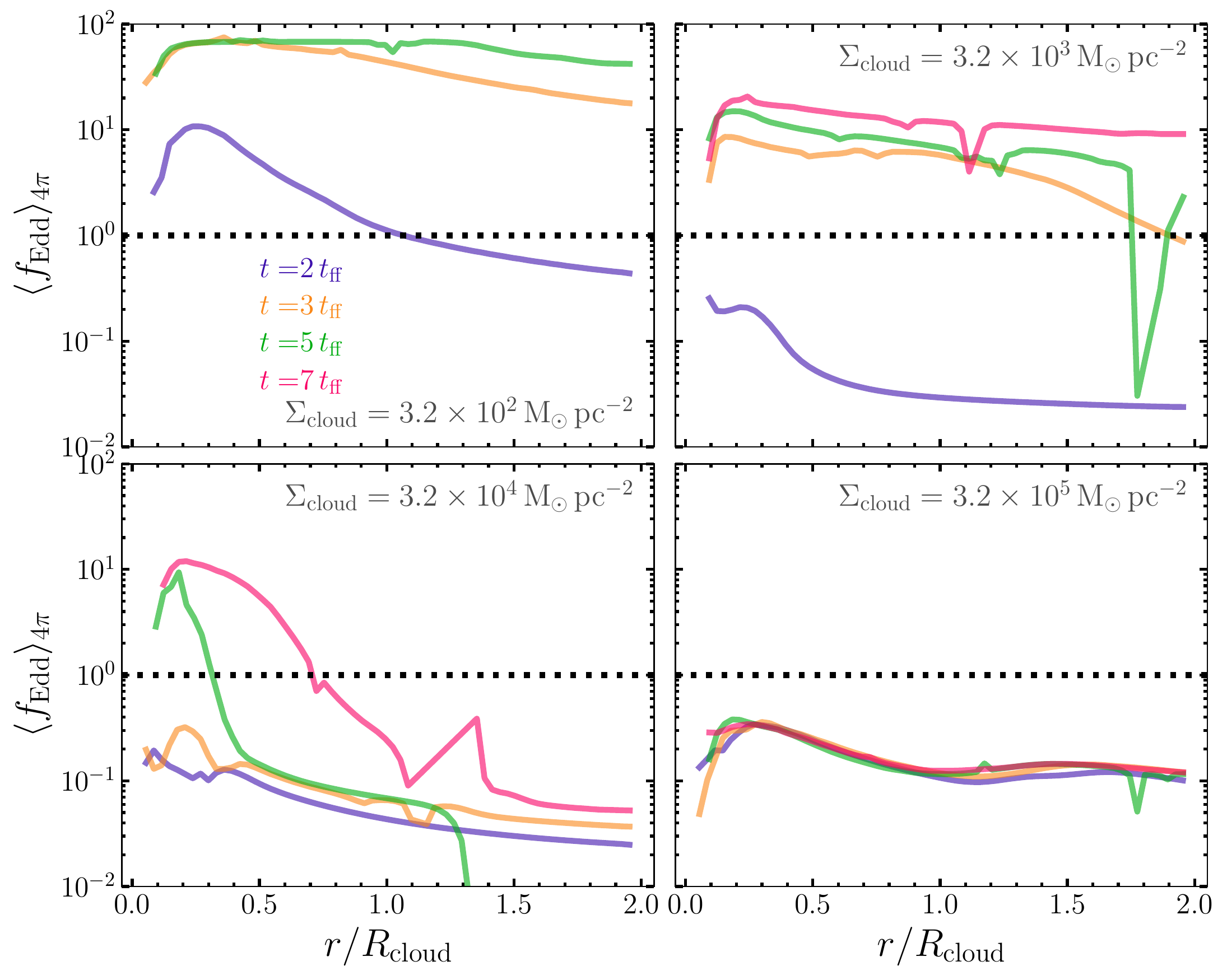}
    \caption{Angle-averaged, volume-weighted Eddington ratio (based on Eq.~\ref{eq:fEdd}) compared at different times for different $\Sigmacloud$ (panels). The corresponding line for $\Sigmacloud = 3.2 \times 10^2 \, \Msolpc$ at $t=7 \, t_\mathrm{ff}$ is not plotted as there is no gas remaining in the domain.}
    \label{fig:CompareEdd}
\end{figure*}

\begin{figure*}
    \centering
    \includegraphics[width = 0.94 \textwidth]{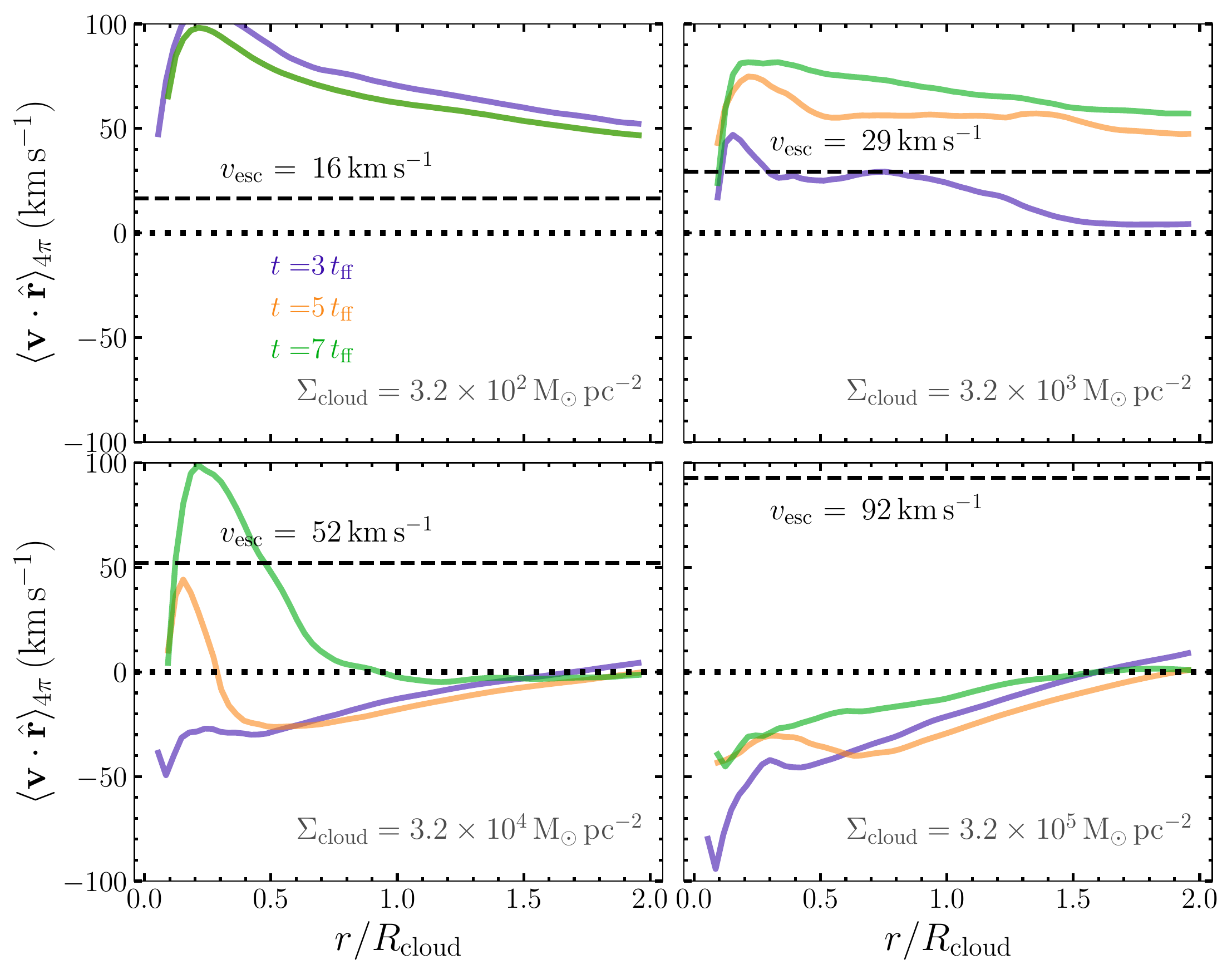}
    \caption{Volume-weighted radial velocity averaged over radial shells at radius $r$ and times $t = [3,5,7] \, t_{\mathrm{ff}}$. The dotted lines indicate zero radial velocities, and the dashed lines indicate the escape speed of the cloud ($v_{\mathrm{esc}}$), with their values annotated.}
    \label{fig:RVelTimes}
\end{figure*}

We discuss the time evolution of our fiducial set of model clouds in this section. The initial turbulent fluctuations form filamentary structures that become gravitationally unstable, and go on to collapse until sink particles (which represent sub-clusters of stars) form. This introduces radiation pressure due to feedback -- i.e., UV photons from the sink particles and the subsequently reprocessed IR photons -- which acts as potential support against gravitational collapse. The subsequent dynamics of the clouds are controlled by whether, and at what point, radiation forces are able to compete with gravity, and therefore depend on $\Sigmacloud$; this can be seen in Figures~\ref{fig:CompareSigmasProj} and \ref{fig:CompareSigmasProjLate}, which show snapshots of the gas surface density at times $t \sim 3 \, t_\mathrm{ff}$ and $t \sim 7 \, t_\mathrm{ff}$ respectively, for the different runs. In model \texttt{S2UVIR} ($\Sigmacloud = 3.2 \times 10^2 \, \Msolpc$), accretion terminates by $\sim 2 \, t_{\mathrm{ff}}$, and radiation forces start driving gas outwards, forming bubbles and filaments characteristic of \ion{H}{II} regions, and evacuating gas from the domain (top-left panel in Figure~\ref{fig:CompareSigmasProj}). Eventually, by $t \sim 4$--$5 \, t_{\mathrm{ff}}$, all the gas is evacuated from the domain, and only the sink particles remain (top-left panel in Figure~\ref{fig:CompareSigmasProjLate}). Model \texttt{S3UVIR} continues to accrete gas even beyond $t \gtrsim 2 t_{\mathrm{ff}}$, and accumulates more mass in sink particles than \texttt{S2UVIR}; however by $t \sim 3 t_{\mathrm{ff}}$, radiation forces become stronger than gravity over a large part of the domain, initiating an outflow (top-right panel in Figure~\ref{fig:CompareSigmasProj}), which becomes stronger and more extended over time (top-right panel in Figure~\ref{fig:CompareSigmasProjLate}). Model \texttt{S4UVIR} evolves similarly at early times, but unlike the earlier cases, there are no signs of radiation-driven outflows at $t \sim 3 t_{\mathrm{ff}}$; once $t\sim 6 \, t_\mathrm{ff}$, however, an outflow is initiated, albeit less pronounced and more asymmetrical than in the cases with lower $\Sigmacloud$ (Figure~\ref{fig:CompareSigmasProjLate}), however showing indications of increasing strength with time. Finally, model \texttt{S5UVIR} continues to collapse for the whole duration of the simulation, with the snapshots showing only signs of infall and rotation (present due to the non-zero angular momentum imparted by the initial turbulent fluctuations), implying that gravity dominates the dynamics in this case.

We quantify the evolutionary stages in the simulations, and the differences with $\Sigmacloud$, by measuring the the star formation efficiency $\epsilon_*$, given by 
\begin{equation}
    \epsilon_* = \frac{M_{*}}{\Mcloud},
    \label{eq:sfe}
\end{equation}
where $M_*$ is final stellar mass, and $\Mcloud$ is the initial cloud mass; Figure~\ref{fig:CompareSFE} shows $\epsilon_*$ (top panel) as a function of time for the different model clouds. We see that the combined gravitational forces from the sink particles and the gas self-gravity increase $\epsilon_*$ for $t \lesssim 2$--$3 \, t_{\mathrm{ff}}$, after which point it saturates at $\epsilon_* \sim 75 \%$ in all runs except the lowest surface density case ($\Sigmacloud = 3.2 \times 10^3 \Msolpc$), which saturates at $\epsilon_* \sim 58 \%$. The former value is similar to that obtained in a control run without feedback (labelled \texttt{NoRT} in Figure~\ref{fig:CompareSFE}). This implies that even though radiation forces in runs \texttt{S3UVIR} and \texttt{S4UVIR} drive outflows, this has no discernible impact on $\epsilon_*$. This is because the outflows begin only after these runs reach their respective final $\epsilon_*$ values. The finding that radiation feedback is unable to regulate $\epsilon_*$ for $\Sigmacloud \gtrsim 10^3 \Msolpc$ is consistent with the results of \citetalias{Menon_2022b}, who only studied the 3 higher $\Sigmacloud$ values in our present simulation suite. We note that although \citetalias{Menon_2022b} only considered the effects of IR radiation pressure, and did not include the UV radiation pressure, \textit{this conclusion remains unchanged}. 

We also quantify the fraction of gas ejected from the volume, $\epsilon_\mathrm{ej}$ (Figure~\ref{fig:CompareSFE}; bottom panel), where 
\begin{equation}
    \epsilon_{\mathrm{ej}} = \frac{M_{\mathrm{ej}}}{\Mcloud},
\end{equation}
such that $M_{\mathrm{ej}}$ is the gas mass ejected from the computational volume. The lowest $\Sigmacloud$ case, as expected, has the vast majority of its cloud mass ejected ($\epsilon_\mathrm{ej} \sim 45 \%$). However, it is more interesting to notice that there are marginal, but non-negligible differences in $\epsilon_\mathrm{ej}$ between run \texttt{S3UVIR} and the higher $\Sigmacloud$/No-RT cases for $t> 4 t_{\mathrm{ff}}$, in spite of their evolution in $\epsilon_*$ being indistinguishable. This is due to the outflows driving mass out of the domain. It is interesting to note that even though the gas morphology and kinematics shows signs of outflowing gas in \texttt{S4UVIR}, the mass removed from the domain is negligible -- as evident from Figure~\ref{fig:CompareSFE}. However, this is likely because the outflows are initiated only at late times, and thus we have not run the simulations for sufficient time for this gas to escape the domain; visual inspection of the time evolution of the clouds confirms this is the case. These results suggest that for i) $\Sigmacloud \sim \mathrm{few} \, 10^2 \Msolpc$, radiation pressure can regulate $\epsilon_*$ and drive a significant fraction of its mass as outflows, ii) for $\Sigmacloud \lesssim 10^3$--$10^5 \Msolpc$, radiation pressure cannot regulate $\epsilon_*$, but once high $\epsilon_*$ is reached, clusters formed in such clouds can drive outflows, and iii) for $\Sigmacloud \gtrsim 10^5 \Msolpc$, radiation pressure can neither regulate $\epsilon_*$ nor otherwise affect the dynamics at any significant level.

To quantify why this is the case, we look at the time-evolution of the Eddington ratio averaged over the full sphere, $\langle f_{\mathrm{Edd}}\rangle_{4\pi}$, where the Eddington ratio $f_{\mathrm{Edd}}$ is given by the ratio of specific radiation ($\dot{p}_{\mathrm{rad}}$) and gravity forces ($\dot{p}_{\mathrm{grav}}$),
\begin{equation}
    f_{\mathrm{Edd}} = \frac{\dot{p}_{\mathrm{rad}}}{\dot{p}_{\mathrm{grav}}}.
    \label{eq:fEdd}
\end{equation}
We use the following procedure to compute $\dot{p}_{\mathrm{rad}}$ and $\dot{p}_{\mathrm{grav}}$. We define a spherical coordinate system centred on the instantaneous centre of mass of the sink particles, and assign every computational cell to one of $128$ radial bins relative to this point. We compute the direction of the radial vector $\hat{\mathbfit{r}}$ relative to the centre of mass, and use it to compute
\begin{equation}
  \dot{p}_{\mathrm{rad}}  = \frac{\left( \kappaRUV \mathbfit{F}_{0,\mathrm{UV} } + \kappaRIR \mathbfit{F}_{0,\mathrm{IR} } \right)}{c} \cdot \hat{\mathbfit{r}},
  \label{eq:forcerad}
\end{equation}
where $\mathbfit{F}_{0,\mathrm{UV}}$ and $\mathbfit{F}_{0,\mathrm{IR}}$ are the radiation fluxes in the co-moving frame of the fluid in the UV and IR band, respectively. The corresponding (specific) gravitational force $\dot{p}_{\mathrm{grav}}$ is given by \begin{equation}
    \dot{p}_{\mathrm{grav}} =  g_{\mathrm{gas}} +  g_{*},
\end{equation}
where $g_{\mathrm{gas}} = -\hat{\mathbfit{r}}\cdot \nabla \Phi_{\mathrm{gas}}$ and $g_{*} = - \hat{\mathbfit{r}}\cdot \nabla \Phi_{*}$, and $\Phi_{\mathrm{gas}}$ and $\Phi_{*}$ are the gravitational potentials of the gas and sink particles, respectively. To compute $\langle f_{\mathrm{Edd}}\rangle_{4\pi}$, we simply take the volume average $f_{\mathrm{Edd}}$ over all the cells in each radial bin.

We plot $\langle f_{\mathrm{Edd}}\rangle_{4\pi}$ for the different $\Sigmacloud$ cases for $t = [2,3,5,7] \, t_{\mathrm{ff}}$ in Figure~\ref{fig:CompareEdd}. We see that the differences and temporal behaviour found in our simulations are consistent with the variations in $\langle f_{\mathrm{Edd}}\rangle_{4\pi}$. The $\Sigmacloud = 3.2 \times 10^2 \Msolpc$ case is super-Eddington at all times for radii $\lesssim \Rcloud$. The $3.2 \times 10^3 \Msolpc$ cloud is sub-Eddington at earlier times ($t/t_{\mathrm{ff}}\lesssim 2$) and then becomes super-Eddington at $t \gtrsim 3 t_{\mathrm{ff}}$. Interestingly, the $\Sigmacloud = 3.2 \times 10^4 \Msolpc$ case -- at late times ($t \gtrsim 5 t_{\mathrm{ff}}$) -- shows a super-Eddington profile for $r \lesssim \Rcloud$, but is sub-Eddington at larger radii, more so at later times. This could potentially explain the behaviour of $\epsilon_{\mathrm{ej}}$ for this run -- i.e., gas at small radii is expelled in an outflow, but rather than escaping to infinity it decelerates and falls back onto the cloud once it reaches larger radii, where the gas is largely sub-Eddington; indeed, this behaviour is visible in the velocity fields in the corresponding panel for this run (lower-left) in Figure~\ref{fig:CompareSigmasProjLate}. Therefore significant mass does not escape the domain in spite of the dynamical signatures of outflows in the gas distributions.

\subsection{Outflows driven by radiation pressure}
\label{sec:Outflows}

Since we find that gas is driven radially outwards by radiation pressure forces in some of our model clouds, in this section we examine the properties of the outflows in more detail. We begin by calculating the radial velocity of the gas $\mathbfit{v} \cdot \hat{\mathbfit{r}}$ over the domain, where $\hat{\mathbfit{r}}$ is the radial unit vector with respect to the centre of mass of the sink particle distribution. We then perform a (volume-)average of this quantity over all solid angles for spherical shells at different $r$ (similar to $\langle f_{\mathrm{Edd}} \rangle_{4\pi}$) to obtain the average radial velocity of gas as a function of radius, i.e., $\langle  \mathbfit{v} \cdot \hat{\mathbfit{r}} \rangle_{4\pi}$. We show this quantity at different times for our fiducial runs in Figure~\ref{fig:RVelTimes}. 

We see that the radial velocities are increasing with time, and are positive over a reasonable extent of the cloud in all cases except the largest $\Sigmacloud$ case, where the gas is inflowing at all radii. In the lowest $\Sigmacloud$ case, the gas is outflowing at up to $\sim 6\times$ the escape speed, even at early times. The \texttt{S3UVIR} run exceeds escape speeds by a factor of \mbox{$\sim 2$--$2.5$} at later times, while the \texttt{S4UVIR} cloud does so only at late times, and even then not over the entire extent of the cloud, consistent with the behaviour of $\langle f_{\mathrm{Edd}}\rangle_{4\pi}$ in Figure~\ref{fig:CompareEdd}. 

\begin{figure}
    \centering
    \includegraphics[width = 0.48 \textwidth]{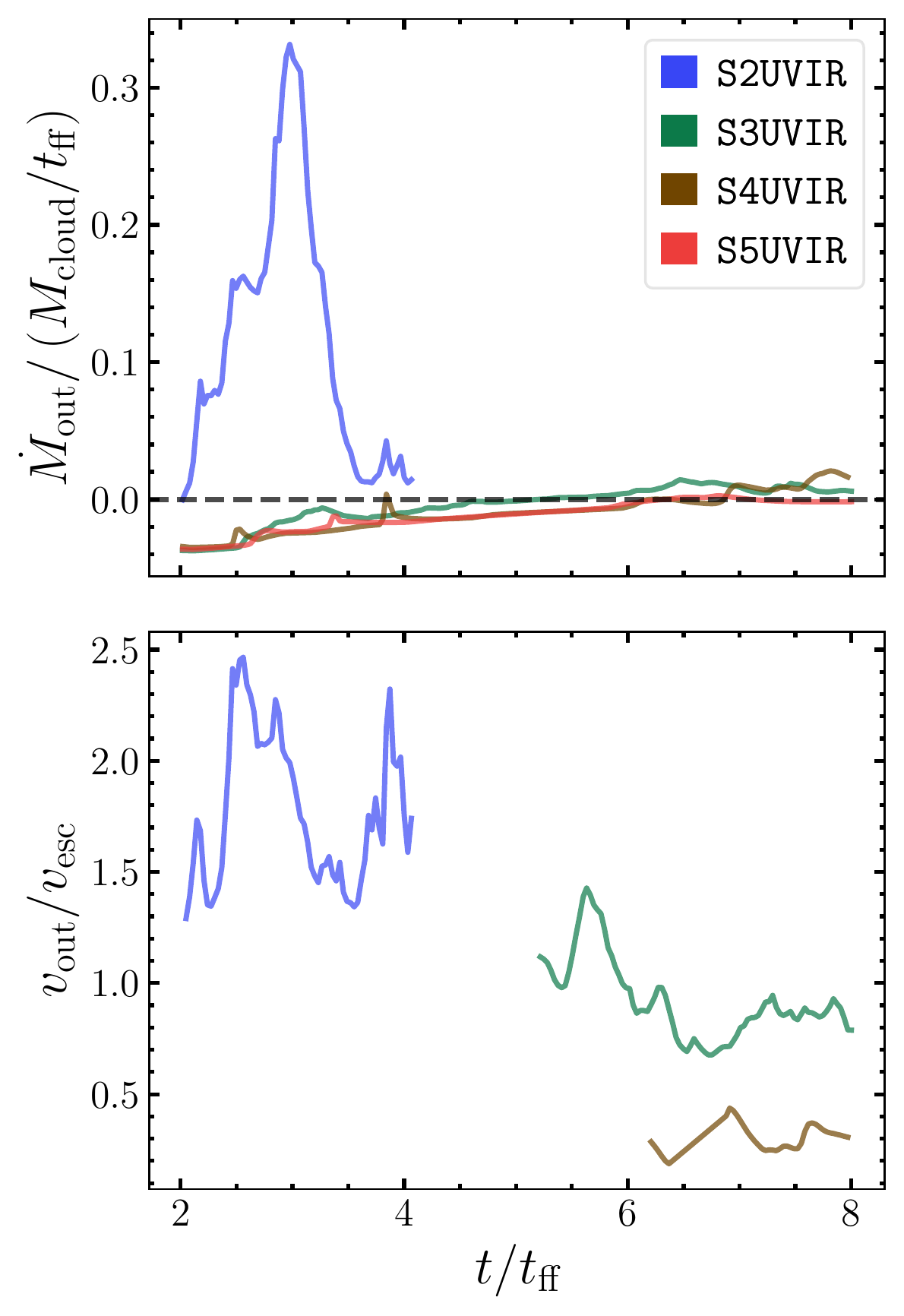}
    \caption{Time evolution of the (normalised) mass outflow rate ($\dot{M}_{\mathrm{out}}$; top), and the momentum-flux weighted outflow velocity ($v_{\mathrm{out}}$), normalised by $v_{\mathrm{esc}}$ (bottom), for runs with different $\Sigmacloud$. We only show $v_{\mathrm{out}}$ for times at which $\dot{M}_{\mathrm{out}}>0$, indicating a net outflow of gas; the corresponding line for \texttt{S5UVIR} is not present as there is no bulk outflow in this case.}
    \label{fig:Outflow_Rate}
\end{figure}

We also compute the mass flux $\Mout$ across the cloud boundary (i.e., the Cartesian surfaces at $\Rcloud$) as a function of time. To compute $\Mout$, we integrate the radial component of the momentum flux over the cartesian surfaces at $\Rcloud$ (denoted by $\partial S$), i.e.,
\begin{equation}
    \Mout = \int_{\partial S} dA \, \rho \left( \mathbfit{v} \cdot \hat{\mathbfit{n}} \right),
\end{equation}
where $\hat{\mathbfit{n}}$ is the unit vector normal to the Cartesian surface, and $dA$ the surface area. We show the time evolution of $\Mout$ for the fiducial set of simulations in Figure~\ref{fig:Outflow_Rate}, scaled by $\Mcloud/t_{\mathrm{ff}}$. We can see that there is a net outflow of material ($\Mout >0$) for all runs except \texttt{S5UVIR}, with the time at which outflows begin increasing with $\Sigmacloud$. To compute a characteristic outflow speed for each case, we define $v_{\mathrm{out}}$, the momentum-flux weighted radial velocity, which is given by 
\begin{equation}
    v_{\mathrm{out}} = \frac{\int_{\partial S} dA \, \rho \left( \mathbfit{v} \cdot \hat{\mathbfit{n}} \right) \mathcal{H} \left( \mathbfit{v} \cdot \hat{\mathbfit{n}} \right) \left( \mathbfit{v} \cdot \hat{\mathbfit{r}} \right)}{\int_{\partial S} dA \, \rho \left( \mathbfit{v} \cdot \hat{\mathbfit{n}} \right) \mathcal{H} \left( \mathbfit{v} \cdot \hat{\mathbfit{n}} \right)}, 
\end{equation}
where $\mathcal{H}$ is the Heaveside step function. We apply the Heaveside filter to ensure that $v_{\mathrm{out}}$ does not diverge even if there is a mixture of outflowing and inflowing gas at the cloud boundary surface, so that $\Mout$ is nearly zero due to cancellations. However, this also means that $v_{\mathrm{out}}>0$ by construction, even if there is no outflow being driven. For this reason we only compute $v_{\mathrm{out}}$ for times where $\Mout>0$; we show this in the lower panel of Figure~\ref{fig:Outflow_Rate}, scaled by the cloud escape speed ($v_{\mathrm{esc}}$; see Table~\ref{tab:Simulations}). We also compute the time-averaged values of $\Mout$ and $\vout$ for times where outflows are driven, which we report in Table~\ref{tab:Results}. We can see that there is a clear progression of $\vout$ from larger to smaller values for higher $\Sigmacloud$. This essentially occurs because the gravitational potential wells are deeper at higher $\Sigmacloud$, and the resulting Eddington ratios are lower (Figure~\ref{fig:CompareEdd}). 

We also compute the total radial momentum in the ejected outflow, $p_{\mathrm{out}}$, given by 
\begin{equation}
    p_{\mathrm{out}} = \int dt \int_{\partial S} dA \, \rho \left( \mathbfit{v} \cdot \hat{\mathbfit{n}} \right) \mathcal{H} \left( \mathbfit{v} \cdot \hat{\mathbfit{n}} \right) \left( \mathbfit{v} \cdot \hat{\mathbfit{r}} \right).
\end{equation}
We normalise this by the final mass of stars formed, to obtain $p_{\mathrm{out}}/M_*$. This is useful to estimate the possible impact the outflows might have on the larger-scale ISM, and to compare with corresponding estimates made for clouds with lower surface densities in earlier studies \citep[e.g.,][]{Kim_2018}. We report the values of $p_{\mathrm{out}}/M_*$ in Table~\ref{tab:Results}. We see that $p_{\mathrm{out}}/M_*$ is relatively low, and is significantly lower than the typical estimates for supernova feedback \citep[e.g.,][]{Kim_2015, Gentry_2017, Gentry_2019}, suggesting that the radiation pressure-driven outflows are relatively insignificant on larger scales. We note, however, that our simulations lack the ionising UV radiation, which could possibly increase the estimates of $p_{\mathrm{out}}/M_*$, although it is likely to be at most a factor $\sim$ few.

\subsection{UV and IR radiation forces}
\label{sec:UVIRForces}

\begin{figure}
    \centering
    \includegraphics[width = 0.48 \textwidth]{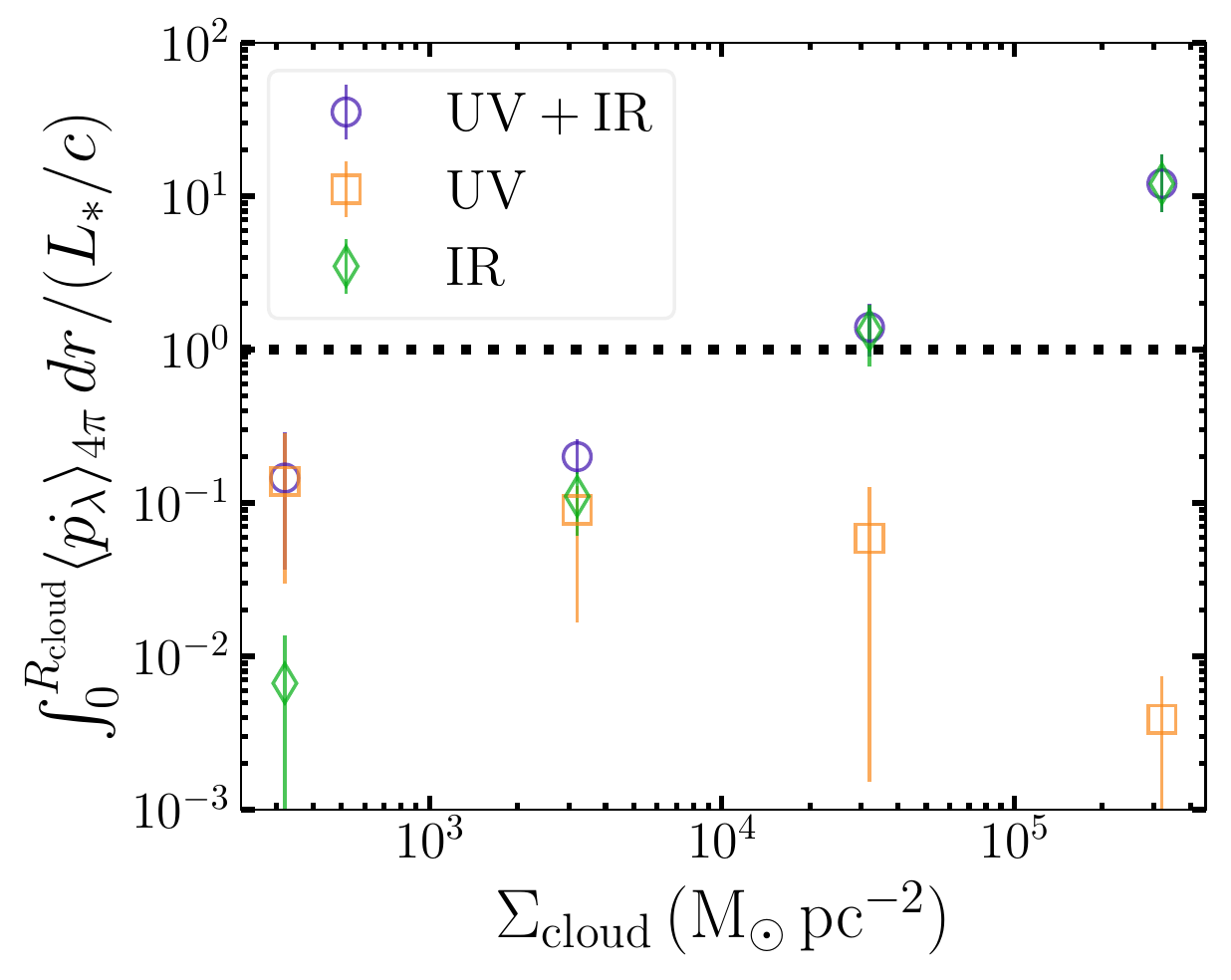}
    \caption{The cumulative momentum rate over all radii scaled by $L_*/c$ that is imparted individually by the UV (Equation~\ref{eq:forceuv}; squares) and IR (Equation~\ref{eq:forceir}; diamonds) radiation pressures, and the combination of the two (circles), in the \texttt{SnUVIR} series.}
    \label{fig:CumForce}
\end{figure}

\begin{figure*}
    \centering
    \includegraphics[width =  0.98 \textwidth]{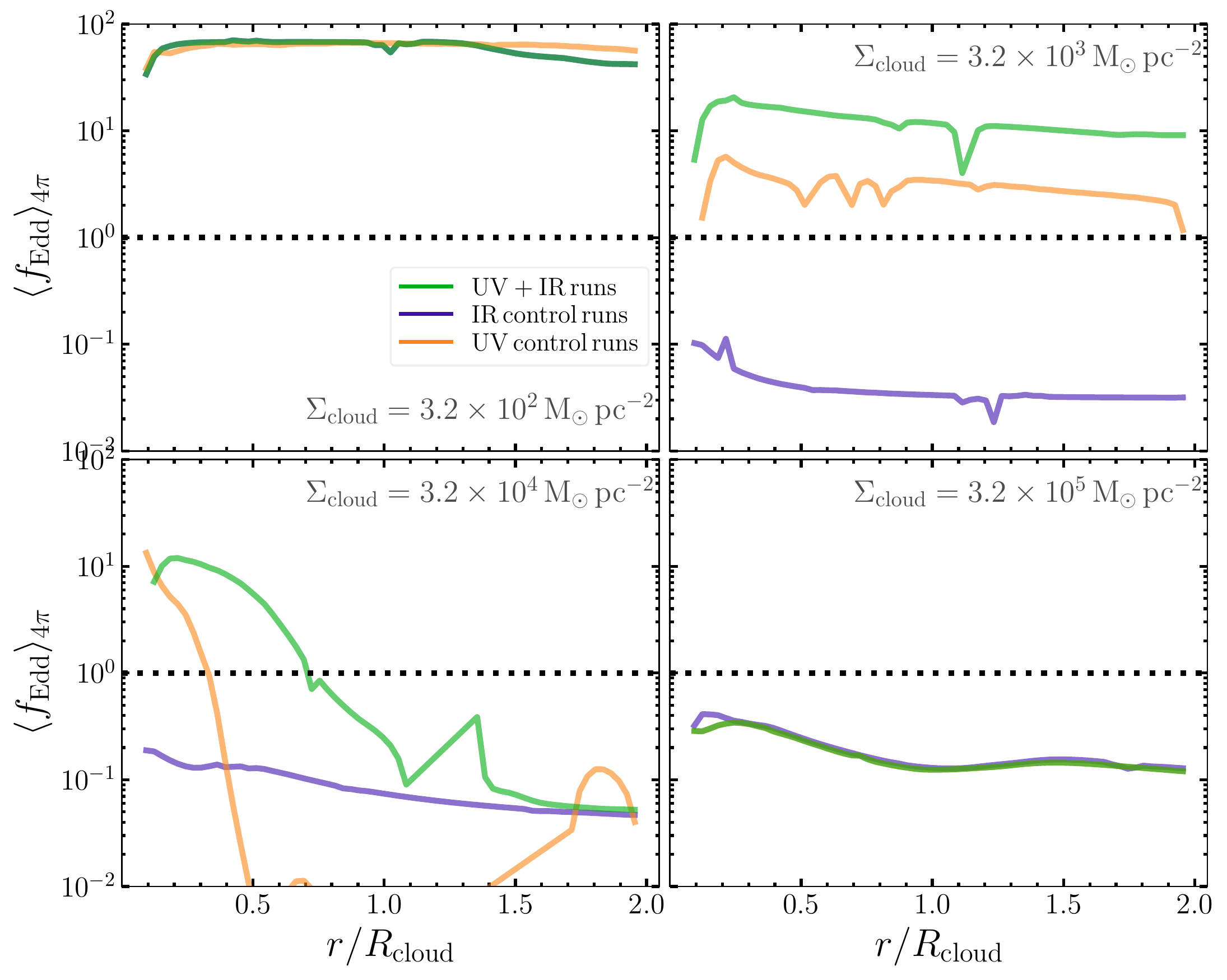}
    \caption{Eddington ratio compared at $t = 7 t_{\mathrm{ff}}$ for different $\Sigmacloud$ (panels) separated by the bands evolved in the simulations.}
    \label{fig:CompareEddUVIR}
\end{figure*}

Our simulations allow us to quantify the relative effects of the radiation forces in the UV and IR band, and thereby their contributions in setting the Eddington ratios in Figure~\ref{fig:CompareEdd}. To do so, we calculate the cumulative radiation pressure forces separately in the UV and IR bands for our fiducial runs. The forces are defined in a similar fashion to Equation~\ref{eq:forcerad}, to produce the cumulative UV radiation force given by 
\begin{equation}
  \dot{p}_{\mathrm{UV},\mathrm{cum}}  = \int_{0}^{\Rcloud} \left( \frac{\kappaRUV \mathbfit{F}_{0,\mathrm{UV}}}{c} \cdot \hat{\mathbfit{r}} \right) \, 4\pi r^2 \, dr
  \label{eq:forceuv},
\end{equation}
and the cumulative IR radiation force
\begin{equation}
  \dot{p}_{\mathrm{IR},\mathrm{cum}}  = \int_{0}^{\Rcloud} \left( \frac{\kappaRIR \mathbfit{F}_{0,\mathrm{IR}}}{c} \cdot \hat{\mathbfit{r}} \right) \, 4\pi r^2 \, dr,
  \label{eq:forceir}.
\end{equation}
In Figure~\ref{fig:CumForce}, we show the time-averaged values of $\dot{p}_{\mathrm{UV},\mathrm{cum}}$ and $\dot{p}_{\mathrm{IR},\mathrm{cum}}$, and their combined force (i.e., $\dot{p}_{\mathrm{UV},\mathrm{cum}} + \dot{p}_{\mathrm{IR},\mathrm{cum}}$), normalised by $L_*/c$, where $L_*$ is the total (UV) luminosity output from the sink particles at a given time. The quantity $L_*/c$ denotes the maximum cumulative momentum that is available in the single-scattering limit -- i.e., this is the maximum possible value of $\dot{p}_{\mathrm{UV},\mathrm{cum}}$. When the cloud is in the multiple scattering limit, the cumulative IR radiation force can exceed this value, and the factor by which it does so is referred to as the trapping factor, $f_{\mathrm{trap}}$. We can see from Figure~\ref{fig:CumForce} that the true cumulative UV radiation force is $\lesssim 0.1 L_*/c$, for reasons that we explore in Section~\ref{sec:radpres_efficiency}. On the other hand, for the IR, $\ftrap \gtrsim 1$ for $\Sigmacloud \gtrsim 10^4 \Msolpc$, with $\ftrap \sim 10$ for the highest $\Sigmacloud$ case. We note that these values of $\ftrap$ are lower than those obtained for the same parameters in the IR-only control runs (2.5 and 18, respectively; c.f.~Figure~19 in \citetalias{Menon_2022b}). This is probably due to \citetalias{Menon_2022b}'s idealised approach of injecting IR photons with a Gaussian source term (Equation~\ref{eq:jstar}), which can lead to a more systematic force in the radial direction than an asymmetric injection of IR photons via the reprocessing of UV radiation. We can also quantify the overall relative importance of the UV and IR radiation forces from Figure~\ref{fig:CumForce}. We can see that the lowest (highest) $\Sigmacloud$ is clearly dominated by the UV (IR) radiation force. The $\Sigmacloud \sim 10^4 \Msolpc$ case is also dominated by the IR radiation force, which is $\sim$ 10 times the UV. On the other hand, for $\Sigmacloud \sim 10^3 \Msolpc$ the forces in the UV and IR bands are comparable, and hence equally important to the dynamics of the clouds. Therefore, this implies that it is important to consider the contribution of \textit{both} UV and IR radiation forces for clouds with $\Sigmacloud \sim 10^3$--$10^4 \, \Msolpc$; however, for clouds that have higher (lower) surface density, the UV (IR) radiation forces are negligible and can be ignored.

\begin{figure*}
    \centering
    \includegraphics[width =  0.98 \textwidth]{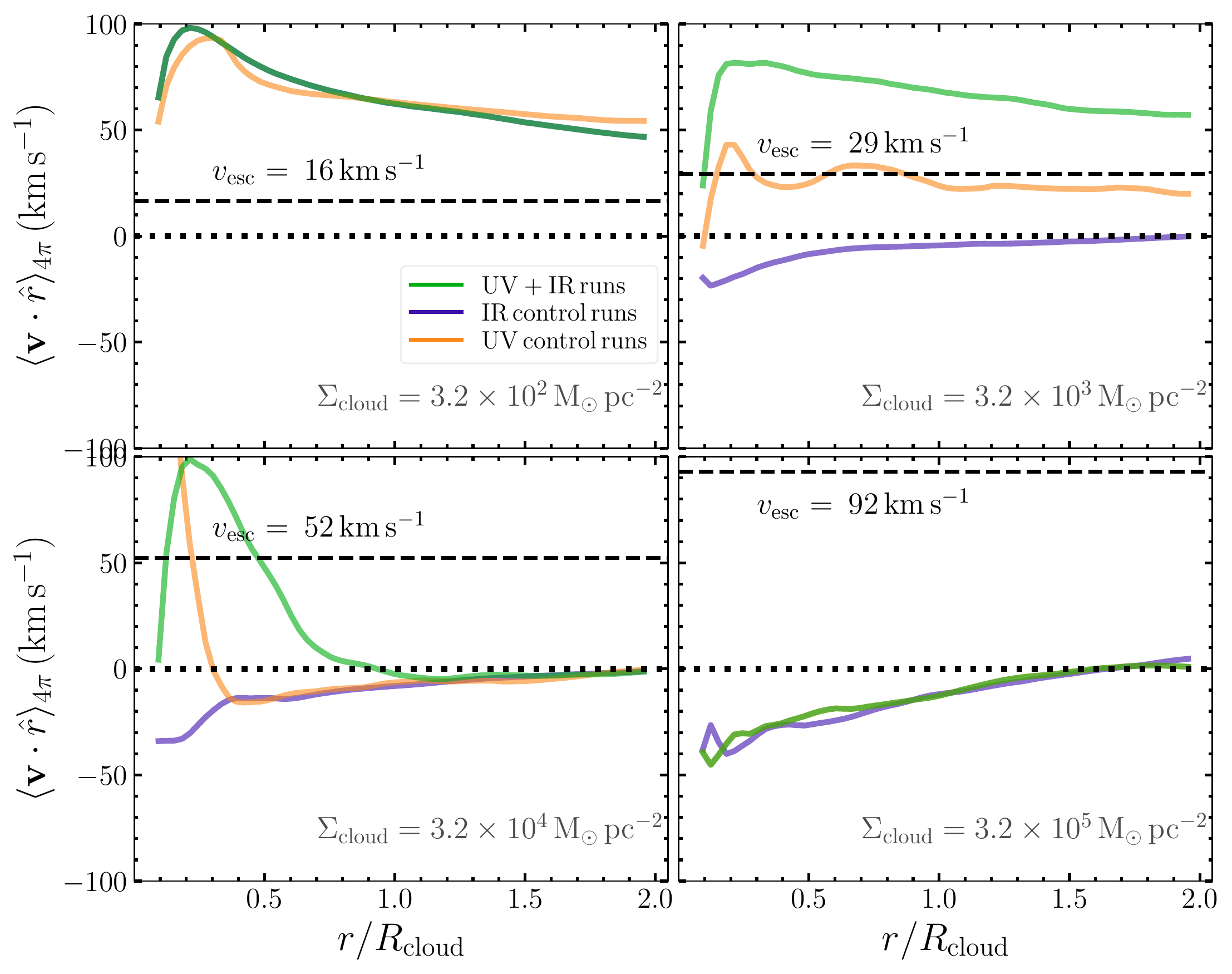}
    \caption{Volume-averaged radial velocity at $t = 7 t_{\mathrm{ff}}$ for different $\Sigmacloud$ (panels) separated by the bands evolved in the simulations.}
    \label{fig:CompareRvelUVIR}
\end{figure*}

Another approach to quantify the relative importance of the UV and IR radiation pressure is to compare the fiducial runs with control runs that do not include one of the bands (i.e., \texttt{SnUV} and \texttt{SnIR} runs; Table~\ref{tab:Simulations}). In Figure~\ref{fig:CompareEddUVIR} we compare $\langle f_{\mathrm{Edd}}\rangle_{4\pi}$ at $t = 7\, t_\mathrm{ff}$ between these simulations. The crucial role played by the UV radiation pressure is clearly visible here; the \texttt{SnIR} runs are all sub-Eddington at all $r$. However, $\langle f_{\mathrm{Edd}}\rangle_{4\pi}$ can be up to factors of a few higher in the UV+IR runs than the UV-only version, especially at smaller $r$. This is likely because the IR radiation pressure is concentrated at small $r$, as the temperatures, and hence the opacities, are lower at larger $r$; visual inspection confirms this is the case. We can also identify the impact the forces have on the dynamics of the clouds by comparing $\mathbfit{v} \cdot \hat{\mathbfit{r}}$ between the runs at the same time, as shown in Figure~\ref{fig:CompareRvelUVIR}. The differences in this quantity between the fiducial runs and the UV/IR-only control runs are quite evident in the cases of intermediate $\Sigmacloud$; the lowest and highest $\Sigmacloud$ cases are more or less indistinguishable from their UV and IR controls runs, respectively, as expected. In both the intermediate $\Sigmacloud$ cases, the UV+IR cases have higher (positive) $\langle v_r \rangle_{4\pi}$ than the UV-only case, and a larger fraction of gas that exceeds the escape speed of the cloud\footnote{The IR-only cases have negative $\langle v_r \rangle_{4\pi}$ at all $\Sigmacloud$, consistent with their sub-Eddington states.}. That being said, even the UV-only cases have radial velocities that exceed $v_{\mathrm{esc}}$, suggesting that outflows are still driven in these runs, but that they involve a smaller fraction of the cloud than in the UV+IR runs.

This behaviour can also be inferred from the time-averaged properties of the outflows driven in the UV-only control runs -- summarised in Table~\ref{tab:Results}. We see that $\vout$ in these runs is more or less comparable to that in the runs with UV+IR for all $\Sigmacloud$. However, for intermediate $\Sigmacloud$, $\Mout$ is lower in the UV-only runs by a factor $\sim 2$, indicating that the inclusion of the IR radiation pressure significantly enhances the mass in the outflows. Similarly, the outflows carry more momentum ($p_{\mathrm{out}}/M_*$) with the inclusion of the IR component, especially for the \texttt{S4UVIR} run. These findings, combined with the behaviour of $\langle f_{\mathrm{Edd}}\rangle_{4\pi}$ and $\langle v_r \rangle_{4\pi}$ in Figures~\ref{fig:CompareEddUVIR} and \ref{fig:CompareRvelUVIR}, suggest that i) the outflows are initiated primarily by the impulse provided by the UV radiation pressure, and ii) the added component of the IR radiation pressure renders a larger fraction of sight-lines around the radiation sources super-Eddington, and thereby entrains more mass into the outflows.

\subsection{Low efficiency of radiation pressure forces}
\label{sec:radpres_efficiency}

\begin{figure*}
    \centering
    \includegraphics[width=0.9 \textwidth]{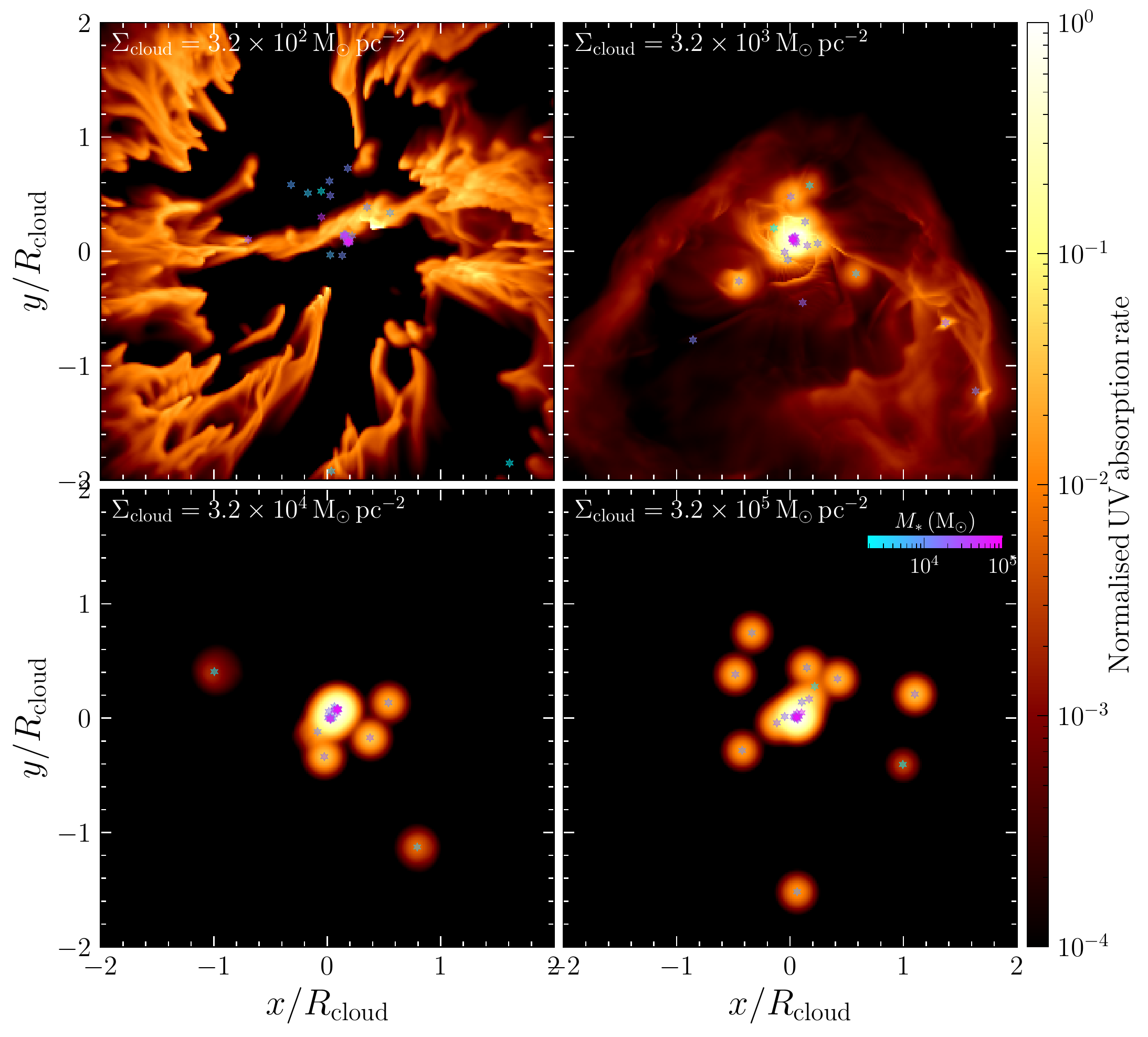}
    \caption{Projected maps of the local UV energy absorption rate at $t = 3 t_{\mathrm{ff}}$ for the different values of $\Sigmacloud$ (panels). The absorption rate is normalised by the maximum value of the quantity in each panel. The star symbols and axes normalisation are similar to those of Figure~\ref{fig:CompareSigmasProj}. We see that the UV radiation is absorbed close to the sink particles for higher $\Sigmacloud$, and the scale over which absorption occurs is small compared to the scale over which the sink particles are distributed. This explains the relatively low efficiency of radial momentum injection by radiation pressure due to cancellation of radiation forces.}
    \label{fig:CompareJstarProj}
\end{figure*}

\begin{figure}
    \centering
    \includegraphics[width=0.48 \textwidth]{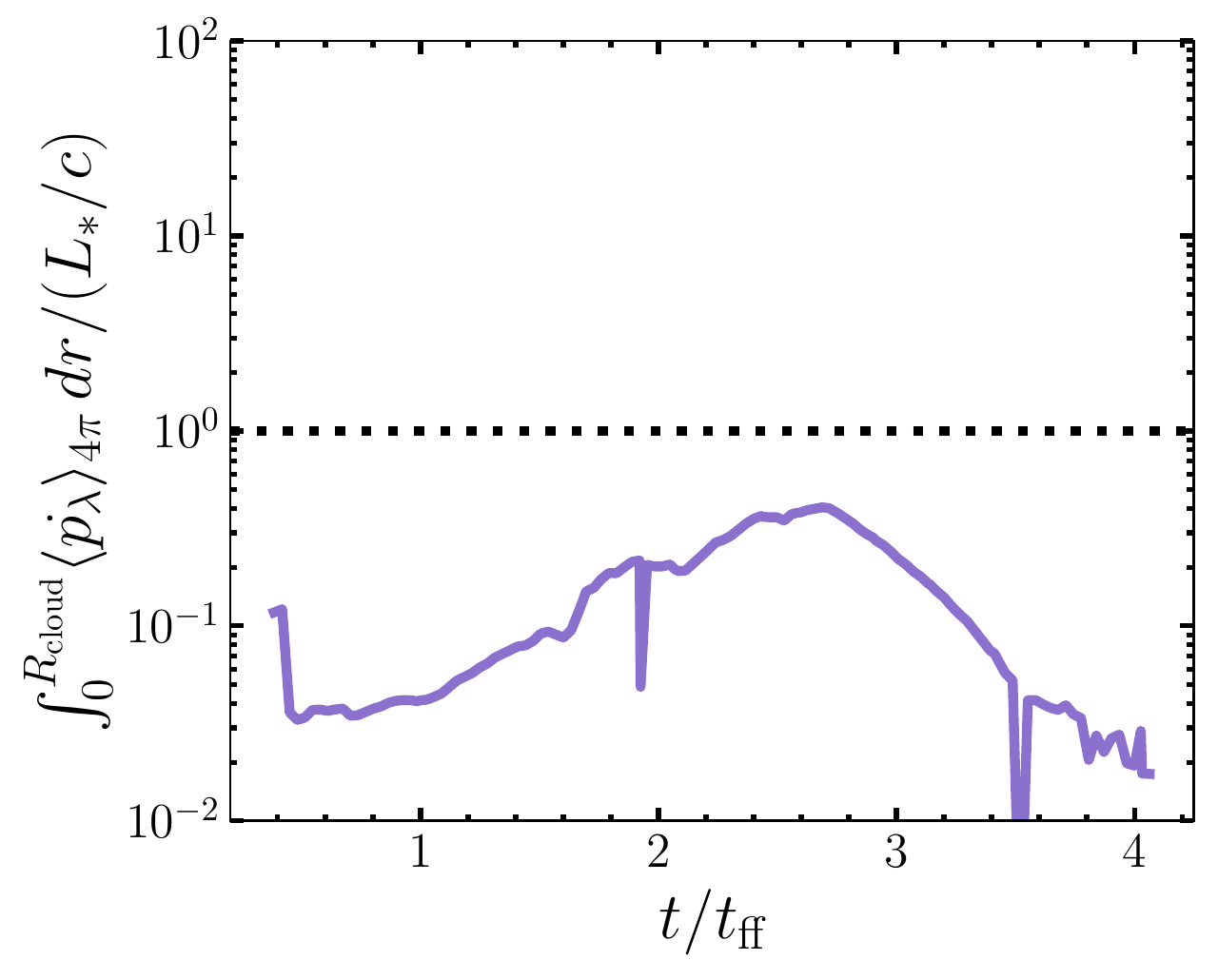}
    \caption{Time evolution of the cumulative UV radiation momentum rate (Equation~\ref{eq:forceuv}, normalised by $L_*/c$ for $\Sigmacloud = 3.2 \times 10^2 \, \Msolpc$ (run \texttt{S2UVIR}) -- similar to Figure~\ref{fig:CumForce}, which shows the time-averaged value. The increase at $t \gtrsim 1.5 t_{\mathrm{ff}}$ corresponds to when radiation pressure drives the UV-absorption region ($d_{\mathrm{UV}}$) outwards, reducing the effects of cancellation due to multiple sources (see text in Section~\ref{sec:radpres_efficiency}); the decrease at $t \gtrsim 3 t_{\mathrm{ff}}$ corresponds to when the outflowing gas opens up channels for radiation to escape the domain.}
    \label{fig:Sigma2_Momentum}
\end{figure}

In Figure~\ref{fig:CumForce}, we quantified the total radial momentum per unit time injected by the UV and IR radiation pressures, in units of $L_*/c$ -- the momentum flux carried by photons from the sink particles. For an idealised spherical distribution with a source at the centre and enough mass around it to be optically thick in the UV, this ratio for the UV case should be 1 (i.e., the momentum per unit time imparted to the gas = $L_*/c$), and should be $\tau_{\mathrm{IR}}$ for IR radiation, where $\tau_{\mathrm{IR}}$ is the cumulative optical depth in the IR. We find that these idealised estimates are much higher than that obtained in our simulations. In \citetalias{Menon_2022b} we explain the origin of this discrepancy for the IR radiation pressure, so we do not repeat that analysis here. However, this still leaves the question of why the cumulative momentum injection rate in the UV $\lesssim 0.1 L_*/c$ in our simulations, as shown in Figure~\ref{fig:CumForce}.

We find that the reason the momentum delivered to the gas is small is due to the cancellation of forces in the radial direction -- with respect to the centre of mass of the sources (sink particles) -- which occurs as most of the UV radiation is absorbed close to the sources, over regions whose sizes are smaller than/comparable to the typical separation between sources. We refer to the scales over which the UV radiation is absorbed and over which the sources are distributed as $d_{\mathrm{UV}}$ and $d_{\mathrm{*}}$, respectively. We can see in Figure~\ref{fig:CompareJstarProj} -- which is a projection of the UV energy absorption rate at $t=5t_{\mathrm{ff}}$ for our fiducial runs -- that $d_{\mathrm{UV}} \lesssim d_{\mathrm{*}}$. For such a situation, the individual (radial) vector forces from each sink, which point radially outwards with respect to the sink, need not necessarily point radially outwards with respect to the centre of mass, leading to a reduction in the radial momentum injection to the cloud. If, on the other hand, $d_{\mathrm{UV}}\gg d_{\mathrm{*}}$, the sinks would all lie within their respective UV absorption zones, and would all contribute positively to the radial momentum.

This helps explain why the efficiency of UV momentum injection is low in all our runs, and more so in the highest $\Sigmacloud$ case ($\sim 0.01 L_*/c$) -- since $d_{\mathrm{*}}$ is very small at these high surface densities (see Figure~\ref{fig:CompareJstarProj}). However, the low efficiency of the $\Sigmacloud = 3.2 \times 10^2 \Msolpc$ case needs further explanation. The cloud is being dispersed by (UV) radiation pressure in this case, and thus $d_{\mathrm{UV}}$ should increase as time progresses, rendering the UV momentum injection more efficient. However, we find that this is countered by another effect: as the cloud expands, this opens up channels through which UV photons escape, decreasing the efficiency of momentum injection, eventually driving it to zero as the cloud is entirely dispersed. It is possible that the combination of these two effects leads to the low time-averaged efficiency of $\sim 0.1 \%$ we find. To investigate whether this is the case, we show the time evolution of the radial momentum injection rate for this run in Figure~\ref{fig:Sigma2_Momentum}. Consistent with our hypothesis, we find that the efficiency is low at early times, then goes up as the bulk of the gas is pushed outwards and the gas distribution increasingly satisfies the condition $d_{\mathrm{UV}}/d_{\mathrm{*}} \gg 1 $. However, for $t > 3 \, t_{\mathrm{ff}}$, the efficiency decreases again due to the escape of UV photons through channels opened up by the dispersing cloud -- the top-left panel of Figure~\ref{fig:CompareJstarProj} provides a visual confirmation of this scenario. By comparing with Figure~\ref{fig:CompareSFE} we also see that i) the increase in momentum injection at $t\gtrsim 1.5 t_{\mathrm{ff}}$ corresponds to when $\epsilon_*$ starts to saturate due to radiation pressure forces and the associated expansion of a shell, and ii) the decrease in momentum injection for $t\gtrsim 3 t_{\mathrm{ff}}$ corresponds to when $\epsilon_{\mathrm{ej}} >0$, indicating that gas has started to escape the domain, opening up channels for UV radiation to escape. 

The aforementioned scenario shows that it matters where the UV photons are absorbed with respect to the distribution of the radiating source(s). An interesting implication of this is that the UV radiation pressure is likely to be a much more efficient feedback mechanism for a single massive star/binary system than for a larger system such as a molecular cloud/star cluster\footnote{Efficient in this context is in terms of the fraction of the total available UV radiation momentum ($L_*/c$) that is effectively imparted to gas in the radially outward direction.}. In the former case, there is less potential for cancellation due to a lower number of sources. In addition, for a massive star, the UV absorption front can be moved outwards due to the destruction of dust; indeed, for a single massive star or close binary, the dust destruction radius is much larger than the system scale, while for even the most compact star clusters the opposite is the case. This further reinforces the point made by \citet{Krumholz_2018} that calculations of radiation pressure feedback are only reliable if they resolve the region over which radiation is absorbed, and that naive subgrid models that do not include effects such as cancellation or the trapping of radiation momentum by gravity on small scales may be unreliable.

A final implication is that any other feedback mechanism that moves $d_{\mathrm{UV}}$ to larger scales -- such as hot stellar wind-driven bubbles or hard ionising radiation that can destroy dust grains and/or provide additional thermal pressure-driven expansion -- would also increase the momentum injected by UV photons closer to $L_*/c$. Therefore, it is possible that the UV momentum injection efficiency is higher if additional feedback mechanisms are active.

\begin{table*}
\caption{Summary of key simulation results.}
\centering
\label{tab:Results}
\begin{threeparttable}
\begin{tabular}{l c c c c c c c c c c}
\toprule
\multicolumn{1}{l}{Model}& \multicolumn{1}{c}{$\epsilon_*$}& \multicolumn{1}{c}{$\epsilon_{\mathrm{ej}}$}& \multicolumn{1}{c}{$\dot{M}_{\mathrm{out}}$}& \multicolumn{1}{c}{$\dot{M}_{\mathrm{out}}/(M_{\mathrm{cloud}}/t_{\mathrm{ff}})$}& \multicolumn{1}{c}{$v_{\mathrm{out}}$}& \multicolumn{1}{c}{$v_{\mathrm{out}}/v_{\mathrm{esc}}$}& \multicolumn{1}{c}{$p_{\mathrm{out}}/M_{*}$}& \multicolumn{1}{c}{$\dot{p}_{\mathrm{cum},\mathrm{UV}}/(L_*/c)$}& \multicolumn{1}{c}{$\dot{p}_{\mathrm{cum},\mathrm{IR}}/(L_*/c)$}& \multicolumn{1}{c}{$\dot{p}_{\mathrm{cum},\mathrm{IR}}/\dot{p}_{\mathrm{cum},\mathrm{UV}}$}\\
& & & $\left[ \Msun \, \mathrm{yr}^{-1} \right]$& & $\left[ \mathrm{km} \, \mathrm{s}^{-1} \right]$& & $\left[ \mathrm{km} \, \mathrm{s}^{-1} \right]$& & & \\
\midrule
\texttt{S2UVIR} &$0.58$ &$0.4$ &$0.029$ &$0.085$ &$28$ &$1.7$ &$14.0$ &$0.14$ &$0.0067$ &$0.048$\\
\texttt{S3UVIR} &$0.74$ &$0.18$ &$0.012$ &$0.0064$ &$25$ &$0.88$ &$0.99$ &$0.09$ &$0.11$ &$1.2$\\
\texttt{S4UVIR} &$0.75$ &$0.14$ &$0.093$ &$0.0086$ &$16$ &$0.31$ &$3.1$ &$0.059$ &$1.3$ &$23.0$\\
\texttt{S5UVIR} &$0.73$ &$0.15$ &$-$ &$-$ &$-$ &$-$ &$-$ &$0.0039$ &$12.0$ &$3100.0$\\
\texttt{S2UV} &$0.6$ &$0.32$ &$0.05$ &$0.15$ &$32$ &$2.0$ &$17.0$ &$0.19$ &$0.0$ &$0.0$\\
\texttt{S3UV} &$0.75$ &$0.16$ &$0.0074$ &$0.0039$ &$30$ &$1.3$ &$0.76$ &$0.08$ &$0.0$ &$0.0$\\
\texttt{S4UV} &$0.75$ &$0.15$ &$0.048$ &$0.0045$ &$20$ &$0.39$ &$0.12$ &$0.065$ &$0.0$ &$0.0$\\
\bottomrule
\end{tabular}
\begin{tablenotes}
\small
\item \textbf{Notes}: Columns in order indicate - Model: model name, $\epsilon_*$: fraction of mass in stars, $\epsilon_{\mathrm{ej}}$: fraction of mass ejected from the domain, $\dot{M}_{\mathrm{out}}$: mass outflow rate, $\dot{M}_{\mathrm{out}}/(M_{\mathrm{cloud}}/t_{\mathrm{ff}})$: mass outflow rate scaled by the cloud mass and free fall time, $v_{\mathrm{out}}$: average momentum-flux weighted outflow velocity, $v_{\mathrm{out}}/v_{\mathrm{esc}}$: outflow velocity scaled by the cloud escape speed $v_{\mathrm{esc}}$, $p_{\mathrm{out}}/M_{*}$: momentum per unit stellar mass carried by the outflowing gas, $\dot{p}_{\mathrm{cum},\mathrm{UV}}/(L_*/c)$: cumulative momentum imparted by the UV radiation pressure in units of $L_*/c$, $\dot{p}_{\mathrm{cum},\mathrm{IR}}/(L_*/c)$: cumulative momentum imparted by the IR radiation pressure in units of $L_*/c$, $\dot{p}_{\mathrm{cum},\mathrm{IR}}/\dot{p}_{\mathrm{cum},\mathrm{UV}}$: ratio of total momentum imparted by IR and UV radiation pressures.
\end{tablenotes}
\end{threeparttable}
\end{table*}

\section{Summary and Discussion}
\label{sec:summary}

We conduct 3D radiation hydrodynamic (RHD) simulations of star cluster formation and evolution in massive, dusty, self-gravitating clouds under the influence of direct UV and dust-reprocessed IR radiation pressure. We use the \texttt{VETTAM} RHD module \citep{Menon_2022} -- which employs the variable Eddington tensor (VET) closure -- to track the propagation of both UV and IR photon bands, accounting for the coupling between the bands due to the reprocessing of UV photons to the IR by dust. We explore marginally bound clouds with gas surface densities of $\Sigmacloud \sim 10^2$--$10^5 \Msolpc$, which ranges from the upper end of the single-scattering limit deep into the multiple-scattering regime (see Table~\ref{tab:Simulations}). We also explore the relative importance of the UV and IR radiation pressure mechanisms by comparing with control runs where one band or the other is omitted. Combining IR and UV radiation pressure, we draw the following conclusions: 
\begin{itemize}
    \item The star formation efficiency $\epsilon_*$ cannot be regulated by radiation pressure for clouds with $\Sigmacloud \gtrsim 10^3 \Msolpc$, even with the inclusion of the UV radiation pressure. In the simulations studied here, which do not include other forms of feedback except radiation pressure, and with isolated clouds that do not receive any energy input from a larger galactic environment, $\epsilon_*$ reaches $\sim 80 \%$ within $t\sim 3 t_{\mathrm{ff}}$ regardless of whether we include IR radiation, UV radiation, or both. We refer the reader to \citetalias{Menon_2022b} (Section 4.4) for a discussion of how these values of $\epsilon_*$ compare to observed estimates.
    \item However, clouds with $\Sigmacloud \lesssim 10^5 \Msolpc$, on attaining high $\epsilon_*$, become super-Eddington and launch radiation-pressure driven radial outflows -- unlike the lack of any dynamical impact of feedback in \citetalias{Menon_2022b} based on IR only. 
    \item The outflows can reach significant radial velocities with high fractions of the escape speed of the clouds, $\vout \sim 0.5$--$2 \, v_{\mathrm{esc}}$ (see Table~\ref{tab:Results}) -- corresponding to $\sim 15$--$30 \, \kms$ -- with the outflow velocity decreasing with $\Sigmacloud$. However, the momentum carried in the ejected outflows ($p_{\mathrm{out}}/M_* \lesssim 10 \kms$) is too small to \textit{directly} affect ISM dynamics at kiloparsec-scales and beyond.
    \item The cumulative momentum imparted by the UV and IR radiation pressure is comparable for $\Sigmacloud \sim 10^3 \Msolpc$, and is dominated by the IR (UV) component in clouds with higher (lower) surface densities.
    \item The characteristic outflow velocity for clouds in the multiple-scattering limit ($\Sigmacloud \gtrsim 10^3 \Msolpc$) does not depend on whether we include only UV radiation pressure or both UV and IR (Table~\ref{tab:Simulations}), but the mass outflow rates and momentum fluxes do: omitting the IR lowers both by factors $\sim 2$. This suggests that the impulse of the UV radiation force provides the launching mechanism of the outflow, while the effect of IR radiation pressure is to render a larger fraction of the gas unbound, thereby allowing the UV pressure to entrain significantly more mass. 
    \item We find that the cumulative momentum imparted by UV photons can be significantly lower than $L_*/c$, more so at higher $\Sigmacloud$. This occurs due to radiation forces cancelling each other out when radiation is absorbed on scales smaller than the typical spatial separation between radiation sources (see Section~\ref{sec:radpres_efficiency}).
\end{itemize}

Our finding that radiation pressure can drive outflows even in clouds with steep gravitational potential wells ($\sim 10^4 \Msolpc$) is interesting and may be significant in the context of the formation and evolution of super-star clusters (SSCs). For instance, \citet{Levy_2021} (\citetalias{Levy_2021} hereafter) analyse high-resolution ($\sim 0.5 \, \mathrm{pc}$) ALMA observations of SSCs in the starburst galaxy NGC~253, and find that a subset of their sample shows signs of (dense-gas) outflows. We can crudely compare the reported properties of the clouds and outflows in their observations (Table~2 in \citetalias{Levy_2021}) with our simulations (Table~\ref{tab:Results}). The clusters with outflows reported in \citetalias{Levy_2021} have $\vout  \sim 6$--$20 \, \kms$, and $p_{\mathrm{out}}/M_* \sim 1$--$5 \, \kms$ -- in reasonable agreement with the values we find. This suggests that radiation pressure is a strong candidate for driving these outflows. Similarly, our results seem to suggest that a potential outflow of molecular gas observed in NGC~2366, coincident with the Mrk~71-A SSC \citep[$\vout  \sim 11 \, \kms$;][]{Oey_2017} is likely driven by radiation pressure.

A minor caveat is that the star clusters with outflows in \citetalias{Levy_2021} have estimated surface densities of $\gtrsim 10^5 \, \Msolpc$, slightly beyond the range of $\Sigmacloud$ where we find outflows are driven. That being said, there are significant uncertainties in the estimated stellar masses and radii of the clusters that go into calculating $\Sigma$ \citep{Leroy_2018}. Moreover, the $\Sigma$ estimated in \citetalias{Levy_2021} is for the observed snapshot wherein the cluster has already formed, whereas $\Sigmacloud$ in our simulations is the initial condition; $\Sigma$ would significantly increase as the cloud collapses under the action of gravity and becomes more compact. We also have to point out larger fractions of our clouds could be ejected at a given $\Sigmacloud$ if i) the clouds were unbound to begin with (i.e., having a larger virial parameter), as suggested by some observations \citep{Evans_2021}, and might be expected to occur in the extreme environments where these clouds form, such as mergers, or ii) through the inclusion of magnetic fields \citep{Federrath_klessen_2012} and/or additional early feedback mechanisms (e.g., stellar winds, photoionisation) in our simulations, and/or iii) a higher dust-to-gas ratio or a more top-heavy IMF, as have been found in some young super-star clusters \citep{Turner_2015}. Therefore, we caution against a direct one-to-one comparison of our model clouds with observed counterparts; rather, we present our findings as evidence that radiation pressure has the momentum budget to drive such outflows. Follow-up observations to better constrain the properties of SSCs and/or extending the sample size would enable testing the viability of these ideas.

While we argue that radiation pressure can indeed launch outflows in star clusters, the same cannot be said for outflows at galactic scales. This is because the galactic discs have significantly larger mass to light ratios than individual young clusters -- they are in the ``old stars'' limit as defined by \citet{Dekel13a} -- and thus the gas within them is sub-Eddington to both single-scattering\footnote{However, local patches within galaxies can be super-Eddington in the single-scattering limit \citep{Thompson_Krumholz_2016, Blackstone_2023}.} \citep{Wibking_2018} and multiple-scattering radiation pressure \citep{Andrews_2011,Crocker_2018a}.
However, outflows launched by star clusters at GMC scales may continue to be accelerated by UV radiation pressure on dust for longer periods, reaching asymptotic velocities of $v_\infty \sim v_{\mathrm{esc}} \left(L_* \kappa_{\mathrm{UV}}/4 \pi GM_*c -1 \right)^{1/2}$, where $L_*$ and $M_*$ are the mass and luminosity of the driving cluster, and $\kappa_{\mathrm{UV}}$ the UV opacity of dust grains, before the wind expands so much that it becomes optically thin and ceases absorbing momentum from the radiation field \citep{Thompson_2015, Raskutti_2017, Krumholz_2017}.
Substituting values adopted in this study for these quantities produces $v_{\infty} \sim 9 v_{\mathrm{esc}}$, which can be up to 500 km/s for the most compact clusters ($\Sigmacloud \gtrsim 10^4 \Msolpc$). This calculation seems to suggest that some gas may be launched to high galactic latitudes by compact star clusters, and potentially even escape the galaxy; that being said, this estimate is highly idealised, and factors such as the ageing of stellar populations, evolution of the mass in the driven shell, and the nature of extended distributions of star formation in the galaxy would significantly affect our estimate. It is also possible that, if there is significant ionising photon escape from the cluster, the asymptotic velocity could be up to several thousand km/s due to the much larger opacity of neutral hydrogen atoms to ionising and Lyman $\alpha$ photons \citep{Komarova21a}. There is scope to explore the longer term evolution of these outflows and their potential observable features in future work.


\section*{Acknowledgements}

We thank the anonymous referee for insightful suggestions that improved the quality of this paper. S.~H.~M would like to thank Todd A.~Thompson, Eve C.~Ostriker, Ahmad Ali, Shane Davis, Jeong-Gyu Kim and Rebecca Levy for insightful discussions and ideas. C.~F.~acknowledges funding provided by the Australian Research Council through Future Fellowship FT180100495, and the Australia-Germany Joint Research Cooperation Scheme (UA-DAAD). M.~R.~K.~acknowledges funding from the Australian Research Council through its \textit{Discovery Projects}, \textit{Future Fellowship}, and Laureate Fellowship funding schemes, awards DP190101258, FT180100375, and FL220100020. We further acknowledge high-performance computing resources provided by the Leibniz Rechenzentrum and the Gauss Centre for Supercomputing (grants~pr32lo, pn73fi, and GCS Large-scale project~22542), and the Australian National Computational Infrastructure (grants~ek9 and~jh2) in the framework of the National Computational Merit Allocation Scheme and the ANU Merit Allocation Scheme.

\textit{Software}: \texttt{PETSc} \citep{PetscConf,PetscRef}, \texttt{NumPy} \citep{numpy}, \texttt{SciPy} \citep{scipy}, \texttt{Matplotlib} \citep{matplotlib}, \texttt{yt} \citep{yt}. This research has made use of NASA's Astrophysics Data System (ADS) Bibliographic Services.

\section*{Data Availability}
Outputs of our simulations would be shared on reasonable request to the corresponding author.



\bibliographystyle{mnras}
\bibliography{RHDCluster,federrath} 


\appendix

\section{Test of the UV Radiation Pressure with \texttt{VETTAM}}
\label{sec:AppendixUVRP}
\begin{figure}
    \centering
    \includegraphics[width = 0.48 \textwidth]{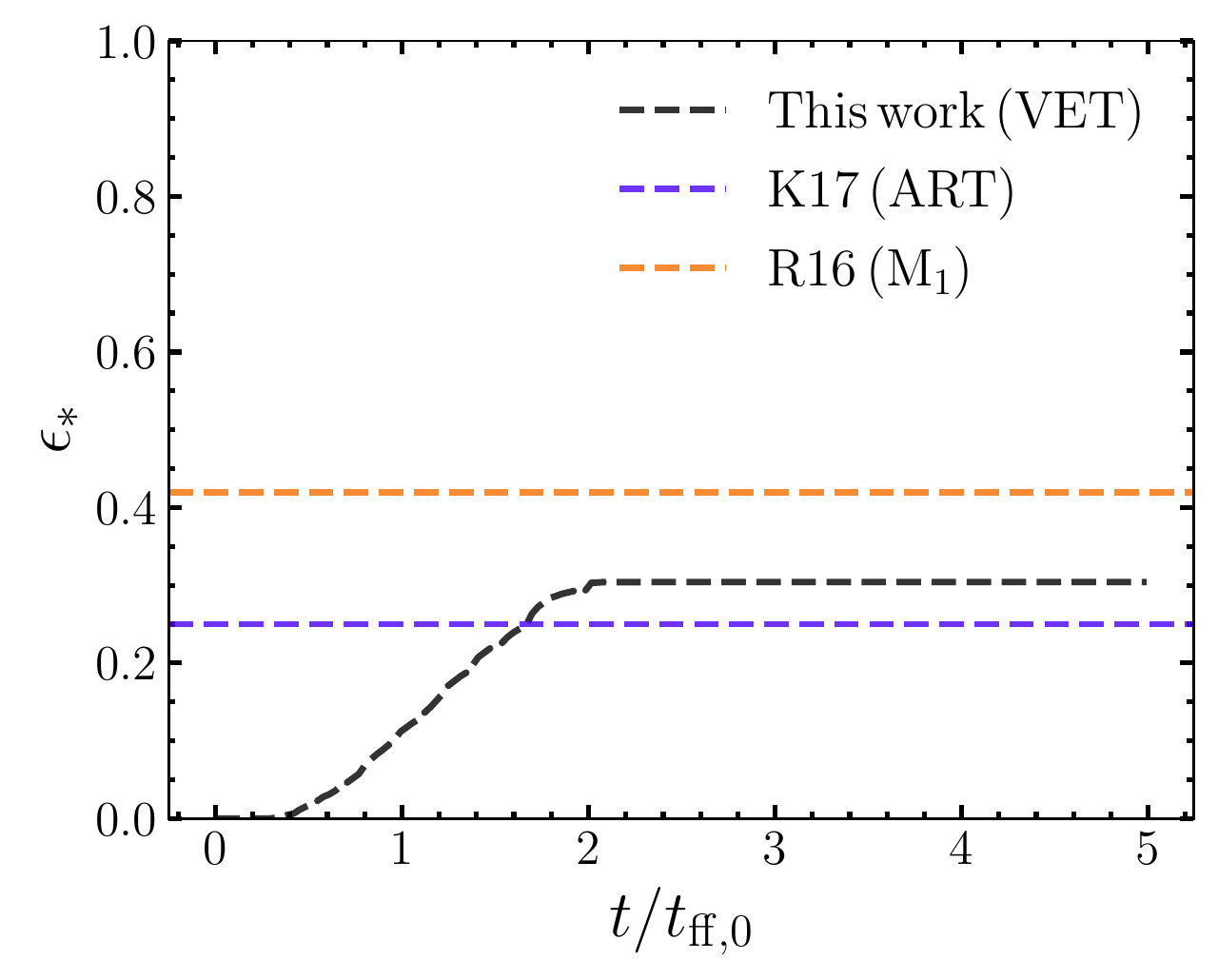}
    \caption{Time evolution of $\epsilon_*$ obtained with \texttt{VETTAM} for the fiducial simulation of \citet{Raskutti_2016}. We also show the final $\epsilon_*$ values obtained in \citet{Raskutti_2016} and \citet{Kim_2017}, which use the $\mathrm{M}_1$ and Adaptive Ray-Tracing (ART) schemes, respectively, to model the propagation of UV photons. We find that the final $\epsilon_*$ obtained with our VET-closed algorithm is reasonably consistent with the ART value, taking into account that $\sim 10\%$ differences in $\epsilon_*$ can be introduced by different random seeds for driving the initial turbulent motions.}
    \label{fig:Appendix_RK16}
\end{figure}

In \citetalias{Menon_2022b}, we compared the outcomes of turbulent star-forming clouds regulated by reprocessed IR radiation pressure obtained with the \texttt{VETTAM} RHD algorithm \citep{Menon_2022}, which uses the VET-closure with that obtained in \citet{Skinner_2015}, which used an $M_1$ closure \citep{Skinner_2013}. We found that the resulting values of the integrated star formation efficiency ($\epsilon_*$) were indistinguishable between the two. However, the reprocessed radiation flux is distributed in a more smooth and isotropic fashion than the direct UV radiation from the stars/clusters, and it is possible that the latter may highlight the limitations of the $M_1$ closure. Indeed, \citet{Kim_2017} repeated the fiducial simulation outlined in \citet{Raskutti_2016} with their Adaptive Ray-Tracing (ART) algorithm, and compared the results to those obtained with the $M_1$ closure used in the original study. They found that the final value of $\epsilon_*$ is lower ($\sim 0.25$) with the more accurate ART method than in the $M_1$ case ($\sim 0.42$). They deduced from the radiation field distributions that the $M_1$ closure underestimates the radiation forces in the vicinity of radiation sources (sink particles), thereby leading to a higher $\epsilon_*$. 

Given this finding, it is interesting to test how our VET-based method performs for this problem; although the VET-closure should be of comparable accuracy to an ART method overall, ART is likely more accurate for the regions in the immediate vicinity of the radiation sources since the moment-based VET method requires some form of ad-hoc injection of photons that is smoothed over some length scale (see Section~\ref{sec:Methods}), and our calculation of the Eddington tensor uses a fixed angular resolution that is in general lower than the angular resolution of an ART method. To test these effects, we repeat the fiducial simulation in \citet{Raskutti_2016} with \texttt{VETTAM}. The model cloud has a value of $\Mcloud = 5 \times 10^4 \, \Msun$, $\Rcloud = 15 \, \mathrm{pc}$, $\alphavir = 2$, and $\sigma_v = 4.16 \, \kms$. The numerical setup is identical to the runs presented in the main part of the paper. The only modification is that we use a light-to-mass ratio of $\psi = 2000 \, \mathrm{erg} \, \mathrm{s}^{-1}$ to match the value used in \citet{Raskutti_2016}. We show the resulting time evolution of $\epsilon_*$ in Figure~\ref{fig:Appendix_RK16}, overplotting the final values of $\epsilon_*$ obtained in \citet{Kim_2017} and \citet{Raskutti_2016}. We find a final value of $\epsilon_* \sim 30\%$, slightly larger than the ART value, but much lower than the $M_1$ case, demonstrating that our VET method can be of comparable accuracy to the ART method for this problem. It is possible that the slightly larger value we obtain is an indication of the unresolved radiation forces in the immediate vicinity of the radiation sources (sink particles) with \texttt{VETTAM}, unlike the ART method. That being said, we found in \citetalias{Menon_2022b} that the turbulent seed used at initialisation can introduce $\sim 10 \%$ differences in the final $\epsilon_*$ \citep[also shown in][]{Skinner_2015}. Accounting for this uncertainty means that our value of $\epsilon_*$ is more or less indistinguishable from the ART value (however, a 10\% uncertainty in $\epsilon_*$ due to the turbulence realisation would not be sufficient to explain the discrepancy with the $M_1$ method discussed above). Therefore, we conclude that a VET algorithm is of comparable accuracy to an ART scheme for modelling point sources in situations where radiation feedback is important. This also serves as a useful test of our algorithm for modelling the direct UV radiation pressure from sink particles.

\section{Convergence Test}
\label{sec:AppendixResolution}

\begin{figure}
    \centering
    \includegraphics[width = 0.48 \textwidth]{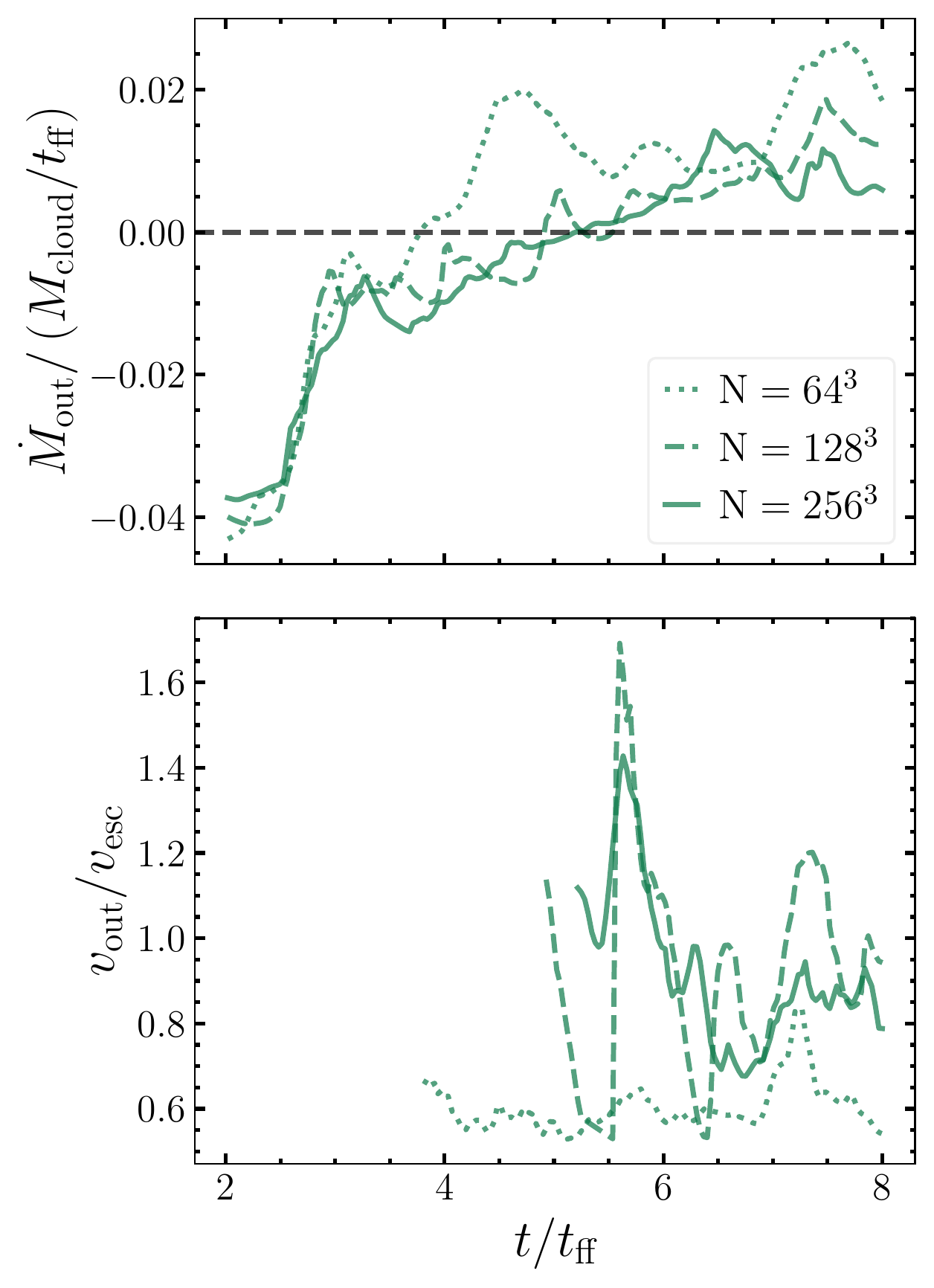}
    \caption{Same as Figure~\ref{fig:Outflow_Rate} for the \texttt{S3UVIR} run, compared for simulations with numerical resolutions of $N=64^3$ (dotted), $N=128^3$ (dashed), and our fiducial choice of $N=256^3$ (solid).}
    \label{fig:Outflow_Appendix}
\end{figure}

We test for numerical convergence of our results by comparing runs with different grid resolutions. We repeat our fiducial simulation, \texttt{S3UVIR}, with uniform-grid resolutions of $64^3$ and $128^3$ to compare with our choice of $256^3$. We found that the obtained values of $\epsilon_*$ and $\epsilon_{\mathrm{ej}}$ were identical to within a few percent, similar to the convergence test presented in \citetalias{Menon_2022b}. Instead, we found it more informative to compare the properties of the radiation-driven outflows. In Figure~\ref{fig:Outflow_Appendix}, we compare the obtained outflow rates and velocities obtained at different resolutions. We can see that the obtained $\Mout$ and $\vout$ are reasonably converged for resolutions of $N>128^3$, with their average values $\lesssim 10\%$ of each other in the $N=128^3$ and $N=256^3$ runs. 

\bsp	
\label{lastpage}
\end{document}